\documentclass[trackchanges,twocolumn]{aastex701}

\usepackage{graphicx}
\usepackage{amsmath}
\usepackage{booktabs}
\usepackage{xcolor}
\usepackage{textgreek}
\usepackage[utf8]{inputenc}
\usepackage[english]{babel}
\usepackage{hyperref}
\hypersetup{unicode,colorlinks=true,linkcolor=linkcolor,citecolor=linkcolor,
            filecolor=linkcolor,urlcolor=linkcolor}
\usepackage{color,colortbl}
\definecolor{linkcolor}{rgb}{0.0,0.3,0.5}
\usepackage{tensind}\tensordelimiter{?}
\DeclareGraphicsExtensions{.bmp,.png,.jpg,.pdf}
\usepackage{verbatim}
\usepackage[normalem]{ulem}
\usepackage{orcidlink}
\usepackage{soul}
\usepackage{tikz}
\usetikzlibrary{arrows.meta,calc,backgrounds}
\newcommand{\fnl}{f_{\rm NL}^{\rm loc}}
\newcommand{\hesp}{\textsc{Hesperos}}
\newcommand{\eosp}{\textsc{Eosphoros}}
\newcommand{\hae}{\hesp~\& \eosp}
\newcommand{\tractor}{\textsc{The Tractor}}
\newcommand{\galsim}{\textsc{GalSim}}

\shorttitle{\hae}
\shortauthors{Cheshire et al.}

\begin{document}

\title{\hae: A Coupled Differentiable Forced-Photometry Operator and Forward Model for SPHEREx Spectrophotometry}

\author[0000-0002-1630-7854]{James~R.~Cheshire~IV}%
\affiliation{Department of Physics, California Institute of Technology, 1200 E. California Boulevard, Pasadena, CA 91125, USA}%
\email[show]{cheshire@caltech.edu}%
\author[0000-0002-3745-2882]{Spencer~Everett}%
\affiliation{Department of Physics, California Institute of Technology, 1200 E. California Boulevard, Pasadena, CA 91125, USA}%
\email{severett@caltech.edu}%
\author[0000-0002-9330-8738]{Richard~M.~Feder}%
\affiliation{University of California at Berkeley, Berkeley, CA 94720, USA}%
\email{rmfeder@berkeley.edu}%
\author[0000-0002-5710-5212]{James~J.~Bock}%
\affiliation{Department of Physics, California Institute of Technology, 1200 E. California Boulevard, Pasadena, CA 91125, USA}%
\affiliation{Jet Propulsion Laboratory, California Institute of Technology, 4800 Oak Grove Drive, Pasadena, CA 91109, USA}%
\email{jjb@astro.caltech.edu}%
\author[0000-0001-5929-4187]{Tzu-Ching~Chang}%
\affiliation{Jet Propulsion Laboratory, California Institute of Technology, 4800 Oak Grove Drive, Pasadena, CA 91109, USA}%
\affiliation{Department of Physics, California Institute of Technology, 1200 E. California Boulevard, Pasadena, CA 91125, USA}%
\email{tzu@caltech.edu}%
\author[0000-0001-6320-261X]{Yi-Kuan~Chiang}%
\affiliation{Academia Sinica Institute of Astronomy and Astrophysics (ASIAA), No. 1, Section 4, Roosevelt Road, Taipei 10617, Taiwan}%
\email{ykchiang@asiaa.sinica.edu.tw}%
\author[0000-0002-4650-8518]{Brendan~P.~Crill}%
\affiliation{Jet Propulsion Laboratory, California Institute of Technology, 4800 Oak Grove Drive, Pasadena, CA 91109, USA}%
\affiliation{Department of Physics, California Institute of Technology, 1200 E. California Boulevard, Pasadena, CA 91125, USA}%
\email{bcrill@jpl.nasa.gov}%
\author[0000-0001-7432-2932]{Olivier~Dor\'{e}}%
\affiliation{Jet Propulsion Laboratory, California Institute of Technology, 4800 Oak Grove Drive, Pasadena, CA 91109, USA}%
\affiliation{Department of Physics, California Institute of Technology, 1200 E. California Boulevard, Pasadena, CA 91125, USA}%
\email{olivier.dore@caltech.edu }%
\author[0000-0002-9382-9832]{Andreas~L.~Faisst}%
\affiliation{IPAC, California Institute of Technology, MC 100-22, 1200 E California Blvd, Pasadena, CA 91125, USA}%
\email{afaisst@caltech.edu}%
\author[0009-0009-1219-5128]{Zhaoyu~Huai}%
\affiliation{Department of Physics, California Institute of Technology, 1200 E. California Boulevard, Pasadena, CA 91125, USA}%
\email{zhuai@caltech.edu}%
\author[0000-0001-5812-1903]{Howard~Hui}%
\affiliation{Department of Physics, California Institute of Technology, 1200 E. California Boulevard, Pasadena, CA 91125, USA}%
\affiliation{Jet Propulsion Laboratory, California Institute of Technology, 4800 Oak Grove Drive, Pasadena, CA 91109, USA}%
\email{hhui@caltech.edu}%
\author[0009-0009-4392-3642]{Gabriela~Torrini}%
\affiliation{IPAC, California Institute of Technology, MC 100-22, 1200 E California Blvd, Pasadena, CA 91125, USA}%
\email{gtorrini@ipac.caltech.edu}%
\author[0000-0001-8253-1451]{Michael~Zemcov}%
\affiliation{School of Physics and Astronomy, Rochester Institute of Technology, 1 Lomb Memorial Dr., Rochester, NY 14623, USA}%
\affiliation{Jet Propulsion Laboratory, California Institute of Technology, 4800 Oak Grove Drive, Pasadena, CA 91109, USA}%
\email{mbzsps@rit.edu}
\collaboration{all}{The SPHEREx Science Team}

\begin{abstract}
We present \hae, a forced-photometry operator and a forward model that, respectively, perform and model the spectrophotometric measurement of the Spectro-Photometer for the History of the Universe, Epoch of Reionization, and Ices Explorer (SPHEREx). Implemented as GPU-accelerated, differentiable JAX programs, the two may be directly composed to measure the response of a downstream product (\emph{e.g.}, the galaxy density field) with respect to a modeled systematic or perturbation through the full measurement pipeline. We describe their design and demonstrate the composition on the response of an astrometric perturbation, and demonstrate how these may be used to generate direct measurement/pipeline response templates $\partial({\rm reconstructed\ density})/\partial({\rm systematic\ amplitude})$ for known image-level effects.\ \hesp\ achieves a factor of $\sim700\times$ increased throughput per node-hour at production scale compared to the prior CPU-based \textsc{Tractor} measurement, previously the limiting step in the experiment's data processing, and propagates automatic derivatives through the full production measurement operator. We validate the measurement operator \hesp\ against the public SPHEREx fork of \tractor, and the forward model \eosp\ against the SPHEREx Sky Simulator and the renders themselves independently against \galsim. Taken together, this represents a powerful increase in capability for scaling, iterating, and optimizing galaxy clustering measurement pipelines.
\end{abstract}

\keywords{Cosmology (343), Large-scale structure of the universe (902), GPU computing (1969), Infrared photometry (792), Astronomy data analysis (1858)}

\section{Introduction}\label{sec:intro}

A key goal of current and upcoming large-scale structure (LSS) surveys, including SPHEREx, is measuring or constraining local primordial non-Gaussianity $\fnl$ through its imprint in the large-scale clustering of galaxies~\citep{dalal2008,dore2014}. The smooth, large-scale features of interest for $\fnl$ are extremely similar in form to the signature of contamination from large-scale modulation of the sample density due to non-cosmological effects, such as anisotropic stellar contamination or the Galactic dust reddening of extragalactic sources. It has been clear since some of the earliest attempts~\citep{pullen2013} that these large-angle systematics and their cleaning have profound effects on the recovery of $\fnl$, and this importance is only growing as the statistical fidelities of cosmological samples continue to improve.

Systematics with measured or modeled spatial profiles, such as stellar contamination and Galactic dust extinction, may have their effects on the density field deprojected directly under some model of how these systematic ``templates'' couple to that density field~\citep{huterer2013,elsner2016,weaverdyck2021}. While some of the obvious systematic contaminants are astrophysical in nature, a significant class enter through the observer- and survey-related effects on the selection of a sample. For SPHEREx, there is some overlap between these, as (for instance) the observing depth is dictated by photon noise from zodiacal emission. The observation-related selection effects may introduce more subtle and nonlinear pathologies to the density field which are difficult to fully correct with template deprojection. For a forced-photometry experiment such as SPHEREx, spurious modulations of the density field may be inherited from both the reference parent catalog's own imaging survey as well as the experiment's own measurement. This paper focuses on the latter class of effect. These observation selection effects are heavily dependent on the photometric measurement itself and its associated uncertainty, which themselves depend on the local background/density, the presence of close/bright blended near-neighbors, point spread-function variation (over the sky and over the focal plane), and exposure-specific transient anomalies or systematics (\emph{e.g.,} cosmic rays, frame-edge ghosts, see~\citealt{fazar2026}). The true selection function of the experiment inherits all of these properties, across all measured pixels, algorithms, and potential software bugs, requiring careful understanding of the measurement operator itself and its coupling to these effects.

A common method for characterizing this response is synthetic source injection, whereby synthetic sources with known properties are inserted directly into real survey images and their responses propagated through the full production pipeline. Evaluating the response against known inputs allows for a significant component of the selection function to be measured directly, at the computational cost of rendering the injected sources and rerunning the production photometry pipeline. The injection procedure makes minimal assumptions about (and captures) nonlinearity present in the response, making it a powerful tool when deployed at scale. \textsc{Balrog}~\citep{suchyta2016,everett2022,anbajagane2025} performs such an analysis for DES, \textsc{SynPipe}~\citep{huang2018} for Hyper Suprime-Cam, and \textsc{Obiwan}~\citep{kong2020} for eBOSS and DESI.\ \citet{kong2025} forward-modeled DESI luminous red galaxy (LRG) density modulations from image simulations with \textsc{Obiwan}; such image simulation-based estimates were used in conjunction with a neural-network systematic regression approach by \citet{rosadomarin2025}, who notably also incorporated a correction for the redshift dependence of the systematics.

Traditional injections are forced to inherit some features of the measurement at face value, notably the point spread functions (PSFs), image-level calibration, astrometry, and on-sky effects such as Galactic extinction, and thus cannot characterize them. If the response of the photometric measurement could be differentiated directly with respect to the amplitude of an image-level systematic or measurement property, and that response is smooth, then the response of the selection to that systematic could be calculated directly, setting the computational cost based only on the measurement operator itself and the desired resolution of the systematic response reconstruction. Differentiable pipelines also facilitate use of gradient-based sampling techniques which may enable other classes of inference, especially when combined with dimension-reduction techniques. Modern array computation frameworks such as JAX~\citep{jax2018github} provide easy-to-use interfaces for constructing pipelines utilizing automatic differentiation (AD), the efficient implementation of which often also comes along with efficient just-in-time (JIT) compilation and graphics processing unit (GPU) acceleration. The construction of a fast, differentiable photometric forward-modeling and measurement pipeline therefore represents a powerful extension in capability in characterizing the selection function, directly complementing source injections.

Characterizing the pipeline response using derivatives has been applied in weak lensing, where the detection responds to the measured shear. This was originally done numerically~\citep{sheldon2017,sheldon2020}, but has also been treated analytically~\citep{li2023} and using AD~\citep{li2024}.\ \citet{mendoza2026} present a fast, differentiable GPU implementation of such a pipeline. These all apply differentiation and gradient-based sampling to the signal of interest, rather than to the systematics. Differentiable, self-optimizing pipelines have been developed and used extensively within high-energy physics~\citep{decastro2019,simpson2023,strong2024,hance2025}, but marginalizing over, or optimizing a pipeline configuration with respect to, a systematic response appears to be an underexplored capability for modern astronomical imaging surveys.

By controlling and composing both an image-level forward model and a forced-photometry measurement pipeline as a single differentiable function of their input properties, we may obtain the image-level photometric measurement significance directly as a function of any variable that those inputs depend on, at the cost of a small multiple of one forward evaluation per differentiated dimension to obtain the automatic derivative. This relies crucially on the performance of the pipeline itself, making optimization essential.

In this work, we present the measurement operator and the forward model (both implemented in JAX), and their composition, which constitute such a pipeline.\ \hesp\footnote{\hae\ refer to the ancient Greek deities embodying the evening and morning stars, respectively. Both refer to the planet Venus, and \hae's names hence reflect how the forced-photometry operator and the forward model represent two aspects of the same measurement.} is a batched, GPU-optimized reimplementation of the SPHEREx windowed forced-photometry measurement, which is validated both against the public fork of \tractor\ \citep{lang2016tractor} and against catalogs produced by the production \textsc{Tractor}-based pipeline from the SPHEREx Science Data Center (SSDC)~\citep{akeson2026}.\ \tractor\ is a general probabilistic photometric measurement forward-modeling utility that may also optimize source positions, shapes and PSFs; SSDC uses it in a mode which fixes these, reducing the problem to a straightforward weighted least-squares problem for each photometric measurement window.\ \hesp\ is designed to tackle this special case directly, making GPU implementation and differentiability practical.\ \eosp\ is an accelerated, differentiable spectrophotometric forward model which renders sources in SPHEREx detector frames, validated against the renders of \galsim~\citep{rowe2015}, and also contains a validated reimplementation of the production binning of raw flux measurements into the ``fiducial'' bins used for redshift fitting (Section~\ref{sec:densmodmaps:fiducial}), and of the redshift estimation stage. Their composition allows the derivative to be taken with respect to a systematic through the rendering and through the same measurement pipeline (\hesp) that may be applied to the real image data.

We note that a differentiable, GPU-accelerated photometry pipeline is not unprecedented: \textsc{AstroPhot}~\citep{stone2023} utilizes accelerated gradient-based optimization of un-forced photometric measurement parameters (positions, shapes, point-spread functions). We instead hold positions and shapes fixed and solve the resulting linear system over the $\mathcal{O}(10^5)$ fitting windows in an exposure, propagating the derivatives through the measurement to the response of a catalog-level quantity.\ \citet{schlafly2018} apply a similar batched linear solve to DECam images using CPUs, though with no derivatives.\ \citet{bahk2026} present a JAX-based reimplementation of \tractor\ and apply it to public SPHEREx data, run on GPUs, though this focuses on depth recovery and external priors on photometry rather than systematics, differentiability, and production throughput.

The map of the components of both and of the paper is shown in Figure~\ref{fig:architecture}. Section~\ref{sec:spxmeas} lays out the SPHEREx measurement problem and the aspects unique to its novel spectrophotometric survey. Section~\ref{sec:hesperos} describes the measurement operator \hesp, and Section~\ref{sec:eosphoros} describes the \eosp\ forward model, detailing design, validation, and throughput. Section~\ref{sec:composition} describes the composition of the two, and establishes the conditions under which the inferred differential responses are trustworthy. Section~\ref{sec:densmodmaps} describes the propagation from measured photometry catalog to measured three-dimensional density fields (implemented within \eosp). Section~\ref{sec:corpus} demonstrates this composition at scale, and the limitations, new capabilities, and relationship to source injection are discussed in Section~\ref{sec:discussion}. We conclude in Section~\ref{sec:conclusions}.

\providecommand{\archtag}[1]{{\scriptsize\textcolor{black!55}{#1}}}
\begin{figure*}
  \centering
  \resizebox{\textwidth}{!}{%
\begin{tikzpicture}[
      x=1cm, y=1cm, font=\footnotesize,
      >={Stealth[length=4pt,width=3pt]},
      box/.style={draw=black!70, rounded corners=1.5pt, align=center, inner sep=3pt,
                  minimum height=1.15cm, fill=white, text width=#1},
      input/.style={draw=black!40, align=right, inner sep=3pt, text width=2.6cm, fill=white},
      chainbox/.style={draw=black!70, rounded corners=1.5pt, align=center, inner sep=2pt,
                       minimum height=0.55cm, fill=white, text width=2.2cm},
      lane/.style={rounded corners=4pt, draw=none, fill=#1},
      lanelabel/.style={anchor=north west, align=left, inner sep=2pt, text=black!75},
      outnode/.style={align=left, inner sep=2pt, text width=2.8cm},
      flow/.style={->, draw=black!75, line width=0.6pt},
      wire/.style={draw=black!75, line width=0.6pt},
      elabel/.style={font=\scriptsize, text=black!60, fill=white, inner sep=1pt, align=center},
      anno/.style={font=\scriptsize\itshape, text=red!60!black, align=center, inner sep=1pt},
      annoarrow/.style={->, draw=red!60!black, line width=0.5pt},
      grad/.style={->, dashed, draw=red!60!black, line width=0.6pt},
      gradtext/.style={font=\scriptsize, text=red!60!black, align=center},
      dot/.style={circle, fill=black!75, inner sep=0pt, minimum size=3.5pt},
    ]
    \hyphenpenalty=10000 \exhyphenpenalty=10000
    \begin{scope}[on background layer]
      \fill[lane=blue!7]    (4.0, 1.1)  rectangle (9.1, 6.8);
      \fill[lane=orange!10] (10.4, 1.3) rectangle (13.2, 6.8);
      \fill[lane=black!8]   (14.0, 0.2) rectangle (16.75, 6.8);
    \end{scope}
    \node[lanelabel] at (4.0, 6.8)  {\eosp\\[-2pt]{\scriptsize (forward model)}};
    \node[lanelabel] at (10.4, 6.8) {\hesp\\[-2pt]{\scriptsize (measurement)}};
    \node[lanelabel] at (14.0, 6.8) {fluxes$\to$density map\\[-2pt]{\scriptsize (built in \eosp)}};

    \node[input, anchor=east] (in_l2)  at (2.6, 5.5)
      {SPHEREx image\\\archtag{pixels, noise, pixel flags}};
    \node[input, anchor=east] (in_tab) at (2.6, 3.5)
      {instrument model\\\archtag{PSF, filters, sky background, observing schedule}};
    \node[input, anchor=east] (in_cat) at (2.6, 1.85)
      {source catalog\\\archtag{positions, shapes, spectra, redshifts}};

    \node[box=3.2cm] (scene) at (6.55, 5.3)
      {simulated observation\\\archtag{sources as SPHEREx sees them}\\
       \archtag{\S\ref{sec:eosphoros:rendering}}};
    \node[box=1.9cm] (direct) at (5.3, 3.1)
      {fast catalog model\\\archtag{fluxes and noise without images}\\
       \archtag{\S\ref{sec:eosphoros:rendering}}};
    \node[box=1.9cm] (render) at (7.8, 3.1)
      {simulated images\\\archtag{sources drawn into pixels}\\
       \archtag{\S\ref{sec:eosphoros:rendering}, \ref{sec:eosphoros:diff}}};

    \node[box=2.6cm] (hesp) at (11.8, 3.55)
      {forced photometry\\\archtag{flux, error, flags per source}\\
       \archtag{\S\ref{sec:hesperos}}};
    \node[outnode, align=center, text width=2.6cm] (l3) at (11.8, 1.85)
      {production-equivalent catalog\\\archtag{\S\ref{sec:hesperos:valcatalog}}};

    \node[chainbox] (psig) at (15.35, 5.2)  {detection significance\\\archtag{\S\ref{sec:composition:psig}}};
    \node[chainbox] (bins) at (15.35, 4.05) {spectral binning\\\archtag{\S\ref{sec:densmodmaps:fiducial}}};
    \node[chainbox] (pz)   at (15.35, 2.9)  {photometric redshifts\\\archtag{\S\ref{sec:densmodmaps:photoz}}};
    \node[chainbox] (tomo) at (15.35, 1.75) {redshift-bin assignment\\\archtag{\S\ref{sec:densmodmaps:tomo}}};
    \node[chainbox] (dens) at (15.35, 0.6)  {3-D density map\\\archtag{\S\ref{sec:densmodmaps:maps}}};

    \node[outnode, anchor=west] (out_nsig) at (16.9, 5.2)
      {response of the number of detected sources to $\alpha$\\
       \archtag{\S\ref{sec:corpus}; Figs.~\ref{fig:response}, \ref{fig:response_lambda}; Tab.~\ref{tab:response}}};
    \node[outnode, anchor=west] (out_dens) at (16.9, 0.6)
      {response of the density map to $\alpha$\\
       \archtag{\S\ref{sec:densmodmaps:maps}, \ref{sec:hesptemplates}; Figs.~\ref{fig:template}, \ref{fig:migration}, \ref{fig:fastvfull}}};

    \draw[flow] (in_cat.east) -- (3.1, 1.85) -- (3.1, 5.0) -- (scene.west |- 0, 5.0);
    \draw[wire] (in_tab.east) -- (3.1, 3.5);
    \node[dot] at (3.1, 3.5) {};
    \draw[flow] ([yshift=-1.5mm]in_l2.east) -- (scene.west |- 0, 5.35);
    \node[elabel] at (3.3, 5.52) {pointing};
    \draw[flow] ([xshift=9mm]in_l2.north) -- ([xshift=9mm]in_l2.north |- 0, 7.15) -- (12.8, 7.15) -- (hesp.north -| 12.8, 0);
    \node[elabel] at (8.0, 7.32) {real pixels, noise, flags};

    \draw[wire] (scene.south) -- (6.55, 4.2);
    \node[dot] (fork) at (6.55, 4.2) {};
    \draw[flow] (fork.center) -| (direct.north);
    \draw[flow] (fork.center) -| (render.north);

    \draw[flow] (render.east) -- (render.east -| hesp.west) node[elabel, midway, above] {simulated\\pixels};
    \draw[flow] (hesp.south) -- (l3.north);
    \draw[flow] (hesp.east |- 0, 3.1) -- (13.6, 3.1) |- (psig.west);
    \draw[flow] ([xshift=8mm]direct.south) -- ([xshift=8mm]direct.south |- 0, 0.9) -- (13.6, 0.9)
                node[elabel, pos=0.62, above] {fluxes, uncertainties} |- (psig.west);
    \draw[flow] (psig.south) -- (bins.north);
    \draw[flow] (bins.south) -- (pz.north);
    \draw[flow] (pz.south)   -- (tomo.north);
    \draw[flow] (tomo.south) -- (dens.north);
    \draw[flow] (psig.east)  -- (out_nsig.west);
    \draw[flow] (dens.east)  -- (out_dens.west);

    \node[anno, anchor=south] (a_flux) at ([xshift=9mm, yshift=4mm]scene.north) {$\alpha$ changes source fluxes};
    \draw[annoarrow] (a_flux.south) -- ([xshift=9mm]scene.north);
    \node[anno, anchor=north] (a_w) at ([yshift=-3mm]in_l2.south) {$\alpha$ changes image noise};
    \draw[annoarrow] (a_w.north) -- (in_l2.south);
    \node[anno, anchor=north] (a_noise) at ([xshift=-3mm, yshift=-3mm]direct.south) {$\alpha$ changes\\predicted noise};
    \draw[annoarrow] (a_noise.north) -- ([xshift=-3mm]direct.south);
    \draw[grad] (4.0, -0.15) -- (19.7, -0.15)
      node[gradtext, midway, below] {derivative with respect to $\alpha$, carried through every stage};
  \end{tikzpicture}}
\caption{Overview of the inputs, outputs, and flow of data for the \hae\ measurement. The left-hand side begins with the input products from the image, instrument, and source catalog, which then provide data downstream as represented by solid arrows. The \eosp\ forward model assembles a scene from the frame geometry, instrument properties, and input source, then either performs a quick catalog-direct evaluation or passes its image-level information to the production measurement operator, \hesp; real data is passed to \hesp\ directly. The resulting catalogs are optionally subjected to a detection-significance step, spectral binning, photometric redshift fitting, and redshift-bin assignment to produce a three-dimensional density map. A systematic amplitude $\alpha$ may enter at the points marked in red; it may either affect the measured flux directly, which must be estimated by rendering the sources and modeling the effect with \eosp, or affect the variance either in the real image or in the fast model's noise prediction. The derivative with respect to the systematic $\alpha$ is propagated through every stage.}
  \label{fig:architecture}
\end{figure*}
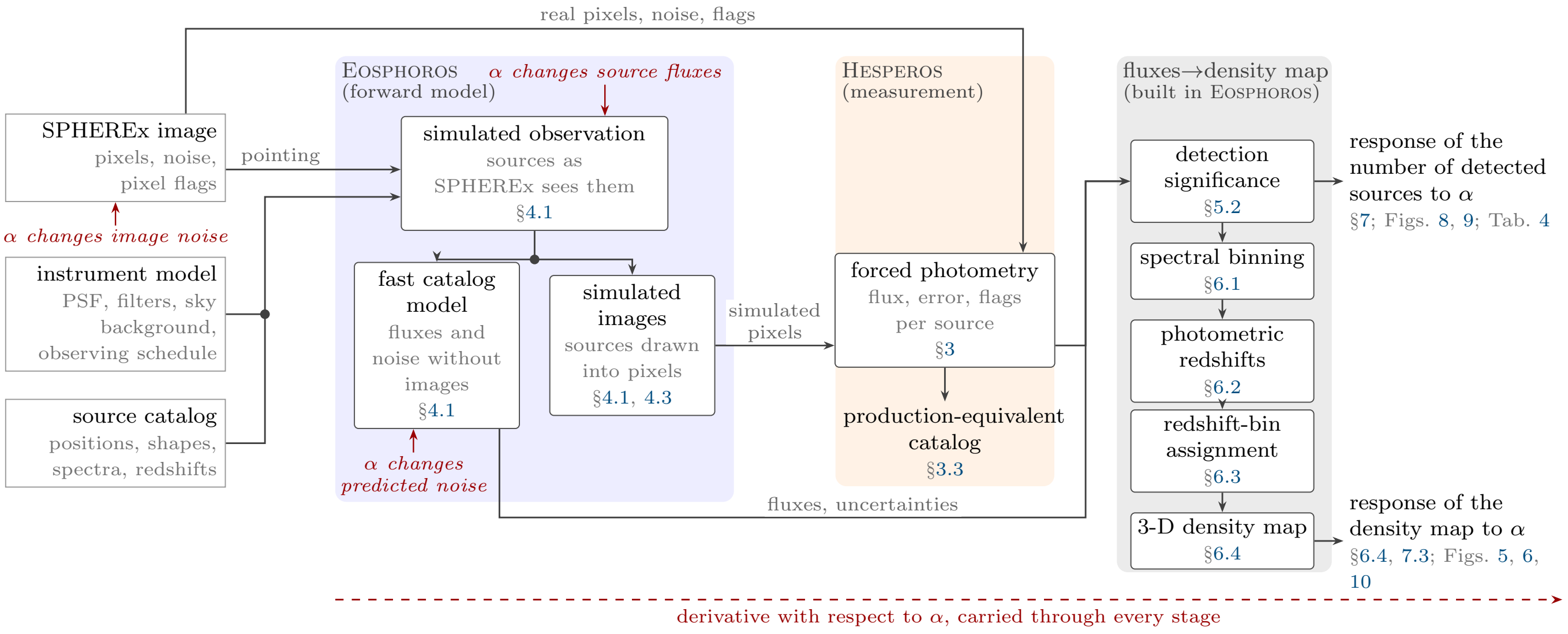

\section{The SPHEREx measurement}\label{sec:spxmeas}

SPHEREx is a NASA Medium Explorer carrying out a novel all-sky spectral survey, enabled by the use of linear variable filters (LVFs) to map the sky in 102 fiducial wavelength channels from $0.75$ to $5.0\,\mu$m~\citep{bock2026}. LVFs vary linearly in passband over one axis of the filter (the image-plane ``$y$'' direction as installed on the six detector arrays comprising the SPHEREx focal plane), enabling an ensemble of measurements of a single source at a variety of detector field-of-view (FoV) positions to reconstruct a low-resolution spectrum ($R \sim 35$--$130$ in 6\farcs15 pixels)~\citep{korngut2026}. The LVF directly couples the act of photometry to the spectral sampling, a regime somewhat familiar to photometric surveys, but distinguished here by the elevated importance of the spectral coupling given the resolution, hence the use of the term ``spectrophotometry''. The wavelength of observation is therefore also a function of geometry, and astrometric perturbations will couple directly to spectral ones, which may directly influence the redshift estimation.

SPHEREx functions as a forced-photometry survey~\citep{lang2016wise,dey2019}, where source positions are defined \emph{ab initio} based on an external reference imaging catalog, the SPHEREx Reference Catalog~(Y. Yang et al. 2026, in preparation). The SPHEREx Reference Catalog is constructed by integrating a range of all-sky and wide-area optical and infrared surveys with curated inputs to enable SPHEREx's core science goals and the construction of its legacy catalog products. The fact that the positions and PSFs (and, for extended object fitting, shapes) are fixed means that the model of the pixel data is linear in the source fluxes, a fact which crucially underlies Section~\ref{sec:hesperos}. An important implication is that these measurements are conditional on having a population of known sources and their positions, along with a known PSF and image calibration; the current implementation is then suitable for a forced-photometry experiment such as SPHEREx, but an extension to source detection using a similar algorithm is planned as future work.

``Levels'' 0--4 are used to refer to various processing stages, referring to both the processing step and the associated output products; see \citet{akeson2026} for details. After initial on-board reduction (Level 0, L0), and calibration/data quality checks (L1), Level 2 (L2) executes photometric and astrometric calibration and pixel flagging to produce calibrated LVF images or spectral data cubes. Level 3 (L3) is the photometric measurement step, and Level 4 (L4) refers to a variety of downstream science analysis pipelines. As a photometric measurement reimplementation and tool to measure the downstream response, \hae\ straddle Levels 3 and 4. We will generally use image frame structures, source positions, variance layers, and flags provided by the Level 2 calibrated image output products, with Level 3 photometry catalogs being targets of comparison rather than products that are used directly.

\section{\hesp: Differentiable forced spectrophotometry}\label{sec:hesperos}

\hesp\ is a differentiable, GPU-batched reimplementation of the \textsc{Tractor}-based SPHEREx Level 3 windowed forced-photometry pipeline described in \citet{akeson2026}. The purpose of \hesp\ is not to optimize the measurement itself relative to the \textsc{Tractor}-based pipeline, but rather to dramatically increase production throughput with the use of GPUs and to enable the differential forward response measurements when composed with \eosp~(Section~\ref{sec:composition}).

\subsection{Design}\label{sec:hesperos:design}

Point-source forced photometry over a blended field performed in isolated windows represents the solution of a large number of small, independent linear systems. Rather than a loop over windows, the photometry measurement may be implemented as a single batched operator that factorizes thousands of independent systems and may be readily adapted to the computing resources in use.

\begin{figure*}
  \centering
  \includegraphics[width=0.6\textwidth]{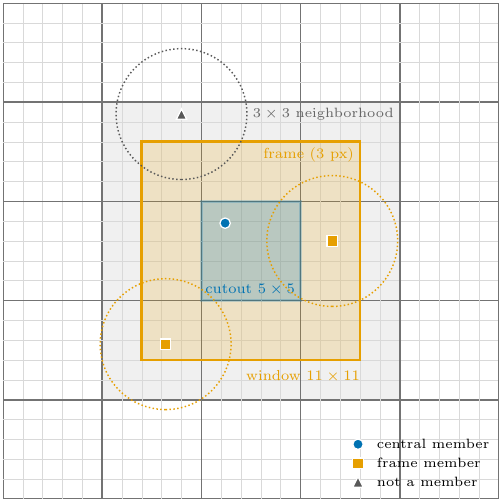}
  \caption{Illustration of fitting window geometry. For each unique cutout in the grid over the image FoV, a window is constructed from the $5 \times 5$-pixel region of that unique cutout (blue) along with a 3-pixel ``frame'' (orange) which accounts for the contributions of neighbors in adjacent cutouts. The $3 \times 3$ neighborhood of cutouts (gray) represents the full set of cutouts over which catalog sources must be considered for inclusion. ``Central members'' which lie in the central unique cutout (blue circles) have their measurements recorded; ``frame members'' in the edge region (orange squares) are measured to understand their contributions to the covariances of central members, but have their measurements discarded; sources outside of the window (gray triangles) are not measured. Dotted circles show the support boundary of a source's ePSF model ($\pm 3.3$-pixels).}\label{fig:windows}
\end{figure*}

An individual SPHEREx Level 2 detector exposure has a resolution of $2048 \times 2048$ $6\farcs15$ pixels, of which a central $2040 \times 2040$ are sensitive to light in the Level 2 images. The active region is divided into a regular grid with cells 5 pixels on a side, giving $408 \times 408 = 166,464$ total windows. Each of these windows is padded by a 3-pixel frame that overlaps its neighbors, yielding an $11 \times 11 = 121$ total window size (the unique cutout plus the frame). A given force-photometered source has its flux, fit uncertainty, background, and quality statistics recorded only for the one window in which it is a member of the unique cutout; it may also be measured in up to eight neighboring windows' frames, for which its measurements would be used to estimate the covariance and then discarded. Because the 3-pixel frame is narrower than the 5-pixel cutout, a given source can only reach the eight adjacent windows, making the membership search an exact $3 \times 3$ neighborhood but potentially truncating contributions from sources which have flux tails beyond this. This windowed setup ensures that the photometry may be cast as a set of fixed linear systems regardless of the catalog, and minimizes redundant observations of the same sources as compared with a per-source centered-cutout approach. The windowed forced-photometry geometry is illustrated in Figure~\ref{fig:windows}. At the detector boundary, pixels beyond the active region are assigned zero weight, to preserve the static dimension of the window, a key performance consideration that will be made use of in Section~\ref{sec:hesperos:perf}.

For a window with $n$ sources, the pixel data $\mathbf{d}$ is linear in the source fluxes,
\begin{equation}\label{eq:linear}
  \mathbf{d} = \mathrm{A}\boldsymbol{\theta} + \boldsymbol{\epsilon},\qquad \boldsymbol{\theta} \equiv (\mathbf{f},\, b),
\end{equation}
where the first $n$ columns of the design matrix $\mathrm{A}$ are the normalized
effective point spread function (ePSF;\ \citealt{anderson2000}) evaluated at each
source's sub-pixel position, multiplying the flux component $\mathbf{f}$ of the parameter vector $\boldsymbol{\theta}$, the final column is constant and carries a locally
flat background of amplitude $b$, and $\boldsymbol{\epsilon}$ is the pixel noise.
The background amplitude is fit jointly with the fluxes, with no prior. The weighted least squares (WLS) solution is then
\begin{equation}\label{eq:wls}
  \hat{\boldsymbol{\theta}} = \mathrm{N}^{-1}\mathrm{A}^\mathrm{T}\mathrm{W}\mathbf{d},
  \quad \mathrm{N} \equiv \mathrm{A}^\mathrm{T}\mathrm{W}\mathrm{A},
\end{equation}
where $\mathrm{W}$ is the diagonal inverse-covariance weight matrix built from the
Level 2 variance plane, $\mathrm{N}$ the normal matrix, and we write $\hat{\mathbf{f}}$ for the source block of
$\hat{\boldsymbol{\theta}}$.

For convenience, we define the naive uncertainty $\sigma_j^{\rm naive} = {\left( \mathrm{N}_{jj} \right)}^{-1/2}$, the diagonal element of the normal matrix which treats a source $j$ as if it were alone in its window, and the marginal uncertainty $\sigma_j^{\rm marg} = {\left[ {\left(\mathrm{N}^{-1} \right)}_{jj} \right]}^{1/2}$, which marginalizes over the covariance with blending neighbor sources and the background. The two are equal only for a source which overlaps with no others, and $\sigma^{\rm marg} \geq \sigma^{\rm naive}$ always. A small ridge is added to the diagonal of $\mathrm{N}$ for degenerate solutions only (solutions that are highly blended and/or have many pixels excluded from the fit due to image quality flags), to regularize them when calculating $\sigma^{\rm marg}$, but the flux is always obtained through Eq.~\ref{eq:wls} without this regularization to avoid biasing it, and these measurements are flagged as degenerate. The full world coordinate system (WCS) including optical distortion is taken into account in the calculation of the pixel solid angle, but this correction is small in any event ($\sim1\%$ over a detector array, $\sim10^{-4}$ over a window).

This setup and the choice to fit the background separately in each window means that the normal matrix of Eq.~\ref{eq:wls} for a full exposure is exactly block-diagonal, so the fit separates exactly and each window reads only the 121 pixels belonging to it. A source $j$ will only be recorded in the single measurement for which it is in the central cutout, independently of other windows which, even if photometering source $j$ as a neighbor, are solved and treated fully independently. This collapses an exposure from a single large linear system into $\sim1.7\times10^5$ small, dense, independent ones. The pixel dimension is made uniform by the window construction and zero-weight edge padding, while the parameter dimension (the dimension over the member sources plus the background term) is made uniform by grouping windows together in ``buckets'' and padding to a small fixed list of source counts with empty, zero-weight sources. Every window in a bucket is therefore independent and has an identical array shape, enabling a dramatic speedup in array performance.

For device memory-specific optimization, each bucket is further divided into ``chunks'' based on a budget of window slots (source positions, real or padded), so that we may scale to a desired memory utilization. Each chunk is issued to the device (the GPU) as a single JIT-compiled program that cuts out each window's pixels, evaluates and normalizes the ePSF, builds and solves every window's linear system at once using a single batched Cholesky factorization~\citep{benoit1924}, and computes the quality flags. Only the results are passed back to the host machine. Source density within an exposure is load-balanced by construction via the buckets/chunks; balancing across images is done by grouping them according to total source count. The quoted throughput numbers average over a representative distribution of source densities.

The $\sim1.7\times10^5$ linear systems then present the host machine with only a small number of device launches of order the chunk count, and the task presented to the GPU is one uniquely suited to it: a composition of simple matrix operations on matrices with static shapes. This architecture is what makes the operation performant by enabling compilability, and also what enables differentiability, since the same computational graph is traversed by the compiler as is traversed by the automatic differentiation system.

The net result is slightly more raw computation than a loop implementation due to the padding step, but deployed more efficiently. On a typical window, the Cholesky solve is about $6\times$ faster than a looped singular-value solve on identical CPU hardware (scaling from $1.2\times$ for a window with $n=10$ members to $\sim 42\times$ at $n=100$), and the solver itself is a small percentage of the total fit time, so the performance described in Section~\ref{sec:hesperos:perf} comes primarily from scale and packing efficiency rather than the solver itself. Derivatives through the Cholesky solve are also stable in nearly degenerate windows, exactly where derivatives through a singular-value solve are unreliable, making the Cholesky approach suitable for a differentiable implementation.

\subsection{Validation against the public \textsc{Tractor} fork}\label{sec:hesperos:valtractor}

We validate against the reference used to drive the SSDC L3 photometry, a public fork\footnote{\url{https://www.github.com/Caltech-IPAC/tractor-spherex/}} of \tractor\ \citep{lang2016tractor}. See \citet{akeson2026} and \citet{huai2026} for details on this fork, but one noteworthy distinction from the upstream \textsc{Tractor} is an updated blending-aware covariance (solving for $\sigma_j^{\rm marg}$ in addition to $\sigma_j^{\rm naive}$). We compare at three different scales: a detailed comparison of a single reference exposure, a comparison of a 20-exposure ensemble spanning a range of source densities against a set of predefined criteria, and a comparison against a subset of a production L3 catalog (Section~\ref{sec:hesperos:valcatalog}).

The fact that the $11 \times 11$-pixel windows are spaced by 5 pixels means that a source is a member of ${(11/5)}^2 \approx 4.8$ windows on average, including the one in which it is measured. The reference exposure used throughout this section has $744,529$ total sources populating $165,076$ out of $166,464$ windows, giving a mean of $21.7$ members per window. This exposure, ID \texttt{2026W12\_2A\_0530\_2D1}, was taken 2026 March 22 and has a field center at right ascension $270.53^\circ$, declination $-20.87^\circ$ on Detector 1 (the shortest-wavelength detector array). This is a dense exposure directly in the Galactic plane ($l = 8.8^\circ,\ b = +0.9^\circ$), representing a deliberately stringent test (see Appendix~\ref{ap:hespval} for examples of windows from this exposure).

Both the \textsc{Tractor} reference and the \hesp\ solution attempt to solve the windowed weighted least-squares problem of Equation~\ref{eq:wls} in 64-bit floating point precision via different algorithms, the former iterative and the latter direct. We compare them in units of the reference marginal uncertainty, $\frac{f_\mathrm{H} - f_\mathrm{T}}{\sigma_\mathrm{T}^{\rm marg}}$, with $f_{\mathrm{H}}$ and $f_{\mathrm{T}}$ being the fluxes estimated by \hesp\ and the \textsc{Tractor} fork, respectively.

\begin{figure}
  \centering
  \includegraphics[width=\linewidth]{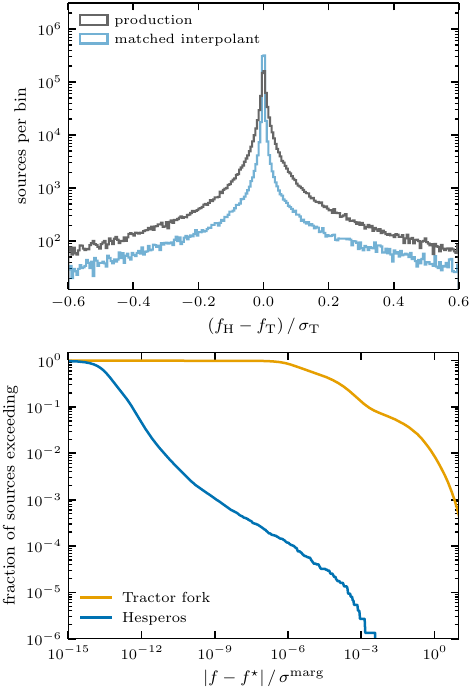}
  \caption{Flux comparison between \hesp\ and the reference \textsc{Tractor} implementation for a matched set of sources in the  reference exposure (disabling either code's fit-quality flags to ensure the same set is being compared on like terms). \textbf{Top}: A histogram of the distance between the two implementations, in units of the reference marginal uncertainty, for the $7.4\times10^5$ matched sources in the reference exposure.  The \hesp\ production interpolant is shown in black, with an interpolant explicitly matched to the reference to isolate only the difference in the solver shown in light blue. \textbf{Bottom}: The cumulative fraction of sources exceeding a given distance from an exact \texttt{numpy} WLS solution $f^\star$, as a function of the solver's own marginal uncertainty. The distribution of \hesp\ solutions is consistently smaller, indicating significantly better agreement with the \texttt{numpy} solution.}\label{fig:pull}
\end{figure}

The chief driver of the differences in flux is the interpolant used to continuously evaluate the tabulated ePSF.\ The SSDC reference uses a FITPACK\footnote{\url{https://perazz.github.io/fitpack/}} spline, versus a cubic B-spline in the \hesp\ implementation. The top panel of Figure~\ref{fig:pull} shows the relative distance between the two implementations' flux solutions for the fiducial \hesp\ configuration for the $744,529$ sources in the reference exposure, as well as a configuration where the \hesp\ solution is held constant by resampling \hesp's ePSF through the reference implementation's spline with the interpolant matched, to highlight the difference in the linear solvers themselves. The agreement is summarized in Table~\ref{tab:validation}. The number of measurements beyond $1\,\sigma^{\rm marg}_{\rm T}$ is 1.4\% with the default interpolant and 0.888\% when the interpolants are matched.

The bottom panel of Figure~\ref{fig:pull} assesses the \textsc{Tractor} reference and the \hesp\ solutions on their own terms, showing the survivorship fractions of both solvers as functions of their own respective marginal uncertainties $\sigma^{\rm marg}$, compared to an (exact, but slow) floating-point 64 \texttt{numpy}~\citep{harris2020} solution for that code's own fit. The reference solution, being iterative, stops short on difficult, highly flagged and blended windows.\ \hesp's direct Cholesky solution on the other hand matches the \texttt{numpy} implementation very closely, with no source beyond $4 \times 10^{-3} \sigma$.

The uncertainty conventions match to a median ratio of 1.000 for $\sigma^{\rm naive}$ and 0.988 for $\sigma^{\rm marg}$, the difference in the latter case arising from \hesp's choice to marginalize over the background term $b$ from Eq.~\ref{eq:linear} in the marginal uncertainty calculation, which the reference implementation chooses to hold fixed. The 20-exposure check passed on 15 of 20, the failing five comprised of three highly flagged/crowded exposures which would be excluded by the analysis, the densest and most crowded exposure of the 20, and one in which a single flag differed due to a boundary effect. This comparison, the uncertainty conventions, and a closer examination of measurements which are highly discrepant between the two implementations is presented in Appendix~\ref{ap:hespval}. 

\begin{deluxetable}{lccc}\tabletypesize{\footnotesize}\setlength{\tabcolsep}{3pt}
\tablecaption{Flux agreement of \hesp\ with the \textsc{Tractor} fork and the production catalog\label{tab:validation}}
\tablehead{
  \colhead{Comparison} & \colhead{Median} & \colhead{99th pct.} & \colhead{99.9th pct.}
}
\startdata
\cutinhead{Reference exposure}
$\mathrm{H}-\mathrm{T}$                          & $0.007$           & $1.4$             & $7.4$            \\
$\mathrm{H}-\mathrm{T}$, matched interpolant     & $6\times10^{-4}$  & $0.87$            & $6.8$            \\
$\mathrm{T}-\mathrm{T}^\star$                    & $2\times10^{-5}$  & $0.84$            & $6.7$            \\
$\mathrm{H}-\mathrm{H}^\star$                    & $4\times10^{-14}$ & $9\times10^{-12}$ & $1\times10^{-9}$ \\
\cutinhead{R7 production catalog}
$\mathrm{H}-\mathrm{L3}$                         & $0.006$           & $0.56$            & $2.5$            \\
\enddata
\tablecomments{Absolute flux differences. H and T are the \hesp\ and \textsc{Tractor} fork fluxes, and $\mathrm{H}^\star$ and $\mathrm{T}^\star$ the exact double-precision solutions for each code's own fit, for the $7.4\times10^5$ sources of the reference exposure of Section~\ref{sec:hesperos:valtractor}. L3 refers to the SSDC production catalog, over $7.2\times10^8$ mutually unflagged measurements (Section~\ref{sec:hesperos:valcatalog}). Code-to-code differences are in units of the \textsc{Tractor} $\sigma^{\rm marg}$, and differences from an exact solution in that code's own $\sigma^{\rm marg}$.}
\end{deluxetable}

\subsection{Validation against the production catalog}\label{sec:hesperos:valcatalog}

The comparison of Section~\ref{sec:hesperos:valtractor} focuses on the solver itself compared against public code on a single exposure. This validates \hesp's flux solution, but does not verify that it reproduces the production catalog with its diffuse background modeling and finalization steps. We therefore also compare against an internally released Level 3 catalog using the Year 1 ($\textrm{R}7.0$) version\footnote{\url{https://irsa.ipac.caltech.edu/data/SPHEREx/docs/SPHEREx_Expsupp.pdf}} of the SSDC L2 and L3 processing software.

For the $7.2 \times 10^8$ measurements that are valid and pass the fit-quality flag in both catalogs, the flux distance in units of the reference's marginal uncertainty $|\Delta f|/\sigma^{\rm marg}$ has a median of 0.006, with a 99th percentile of $0.56\sigma$ and a 99.9th percentile of $2.5\sigma$. Per-exposure, the median signed flux distance has a median magnitude of $7 \times 10^{-5} \sigma$, never exceeding $0.18\sigma$ across the $16{,}362$ exposures with at least 30 valid measurements.

The composition of the output catalog and incorporation of L2 image-level flags is identical. One notable asymmetry is the number of measurements flagged by the fit step itself based on condition number, with \hesp\ flagging approximately $1.5\times10^6$ (0.2\% of the total mutually unflagged population) measurements that were not flagged by the production catalog, with only $5.5\times10^3$ in the other direction. These are driven by near-degenerate blends, and while the two implementations agree closely on which fields are poorly conditioned, they impose thresholds of differing strictness in constructing the catalogs. The flagged fraction is a strong function of Galactic latitude as well, due to the variation in local conditions and density, though this also cautions against the overinterpretation of statistics which aggregate over many regimes.

\subsection{Throughput}\label{sec:hesperos:perf}

Rates are given as raw wall-clock over long work-lists, at each platform's measured packing optimum, including the full startup cost\footnote{Each process pays a substantial initialization cost from the JIT compilation, requiring longer benchmarks to understand production-relevant throughput.}. An exposure is measured by several independent processes running on the same GPU device, allowing computation to continue on the GPU while one process is running its host-side leg. The number of processes needed to saturate the GPU or GPUs was measured and optimized per architecture. Benchmarks here are run over a set of $3{,}674$ Detector 1 exposures consisting of two $1837$-exposure, non-consecutive weeks of observing data ($2.5 \times 10^9$ flux measurements).

A single exposure in a single process in point source mode, warm, runs in about 0.180 seconds on an NVIDIA GH200~\citep{gh200}. The point-source photometry mode sustains $2.97\times10^4$ exposures per hour on that same GH200, and around $8.5\times 10^4$ exposures per hour on a four-GPU GB200~\citep{gb200} node with some higher run-to-run variance. In extended mode, where each source additionally requires a S\'ersic convolution, the GH200 sustains $3.5 \times 10^3$ exposures per hour. This is about $8.5\times$ the point source path, though this depends on the distribution of profiles and density. The same ensemble on the four-GPU GB200 node sustained $1.24 \times 10^4$ exposures per hour, scaling slightly better than the point-source path ($\sim 3.5\times$ versus $\sim 3\times$ the single-GH200 performance) due to the extended path's packing being more memory-dependent and benefiting more from the additional high-bandwidth memory of the Blackwell architecture cards. Six processes were found to saturate the GH200 in point-source mode and four in extended, while the GB200 cards benefited from up to 6 per-card in point-source mode and 8 in extended mode.

Running both the SSDC reference implementation and \hesp\ on identical CPU hardware (dual-socket 128-core AMD EPYC 7763 ``Milan'' with 256 GB of system RAM, the same hardware on which SSDC run the production pipeline) to isolate the algorithmic differences from the batching/hardware acceleration, we find memory to be the limiting factor, with the reference SSDC pipeline processing around 122 exposures/CPU-node-hour with the node fully packed versus around 107 exposures/CPU-node-hour on the same hardware with \hesp. Extensive effort was not taken to optimally pack the CPU \hesp\ configuration on the EPYC hardware; this could be further optimized to some degree, but the salient point is that the vast bulk of the improved performance offered by \hesp\ is driven by the suitability of this problem to GPU hardware, rather than the fitting algorithm itself. The throughput of \hesp\ on a fully packed four-GPU GB200 system represents a $\sim 700 \times$ increase in throughput per unit of compute allocation ($\sim240\times$ for the single-GPU GH200 nodes). A comparison of the performance for the tested configurations is shown in Table~\ref{tab:throughput}.

\begin{deluxetable*}{llcrr}
\tablecaption{\hesp\ throughput compared with the SSDC reference implementation\label{tab:throughput}}
\tablehead{
  \colhead{Code} & \colhead{Hardware per node} & \colhead{Mode} &
  \colhead{Exposures / node-hour} & \colhead{vs.\ SSDC}
}
\startdata
\cutinhead{Benchmark: $3{,}674$ Detector~1 exposures}
SSDC (\tractor\ fork) & $2\times$ EPYC 7763  & point    & 122              & 1       \\
\hesp                 & $2\times$ EPYC 7763  & point    & 107              & 0.88    \\
\hesp                 & $1\times$ GH200              & point    & $2.97\times10^4$ & 240     \\
\hesp                 & $4\times$ GB200              & point    & $8.5\times10^4$  & 700     \\
\hesp                 & $1\times$ GH200              & extended & $3.5\times10^3$  & \nodata \\
\hesp                 & $4\times$ GB200              & extended & $1.24\times10^4$ & \nodata \\
\cutinhead{DR1 production: all six arrays}
\hesp                 & $4\times$ GB200              & point    & $6.9\times10^4$  & \nodata \\
\hesp                 & $4\times$ GB200              & extended & $1.11\times10^4$ & \nodata \\
\enddata
\tablecomments{Raw wall-clock throughput including startup cost. CPU rows are dual-socket 128-core AMD EPYC 7763 nodes; GPU rows are NVIDIA GH200 and four-GPU GB200 nodes. Benchmark rows are at each platform's optimal memory packing. The GB200 point-source rate spans $8.1$--$9.2\times10^4$ across repeated runs. DR1 rows are the full production runs on four GB200 nodes: $1{,}312{,}429$ exposures in 18.9 node-hours (point source), and the $1{,}116{,}233$ exposures containing at least one extended source in 100.3 node-hours (extended). The SSDC extended-source mode is experimental and was not run.}
\end{deluxetable*}

To test the extrapolations and exercise throughput at scale, the entirety of the first-year ``DR1'' data release was reprocessed from the DR1 Level 2 calibrated images using \hesp, run in both point-source and extended modes\footnote{The SSDC extended-source mode is experimental and was not run for DR1, so we defer detailed validation and provide this for cross-comparison of the \hesp\ point-source vs.\ extended-source photometric throughput.}, shown in the bottom two rows of Table~\ref{tab:throughput}. This represents slightly more than the first year of SPHEREx data, including the first two surveys of the full sky: $1{,}312{,}429$ exposures, of which $1{,}116{,}233$ contain at least one source that is designated as extended. Forced photometry is processed for extended sources using S\'ersic indices and types from the DESI Legacy Imaging Survey~\citep{dey2019}; sources without information on extent are processed as point sources. Averaged over the full ensemble of images and catalogs, the performance degrades slightly versus the shorter benchmarks; this is largely driven by additional file loads and other CPU- or filesystem-limited operations. The $\sim1.3$M-exposure ensemble was processed in point-source fitting mode in 18.9 GB node-hours (75.6 Blackwell GPU-hours), and the set of $\sim 1.1$M exposures containing at least one Legacy Survey extended source in 100.3 node-hours (401.6 GPU-hours). Photometry for the entire survey in the default point-source mode is processed in under a node-day for the four-GPU node, or a few days on a single-GPU GH200. Processing a small subset of observations or a specific region of interest can be done in a matter of hours or minutes. Variations of the measurement (such as the ePSF or other calibration factors) may have their impacts on downstream data assessed immediately, automatic derivatives used to probe the response cheaply, and techniques that rely on rerunning the measurement such as injections may be massively accelerated and scaled-up.

\section{\eosp: The forward model}\label{sec:eosphoros}

\eosp\ is a spectrophotometric forward-modeling software that renders sources into detector frames. Validated against the full instrument-based SPHEREx Sky Simulator~\citep{crill2025}, \eosp\ represents a highly-accelerated and more approximate approach which nevertheless incorporates the full spatial-spectral LVF-PSF coupling intrinsic to SPHEREx's unique spectrophotometric measurement.

\subsection{Spectrophotometric rendering}\label{sec:eosphoros:rendering}

Pre-flight in-lab measurements~\citep{hui2026} have characterized the instrument's spectral response for each detector array, including its LVF.\ For a source at a detector position $(x,\ y)$, we retrieve from the table the LVF response $T(\lambda; x,y)$ at that pixel. The in-band flux-density is then the bandpass-weighted mean,
\begin{equation}\label{eq:lvfflux}
  \bar{f}_\nu = \frac{\int f_\nu (\lambda) T(\lambda;x,y) \lambda^p {\rm d}\lambda}{\int T(\lambda;x,y) \lambda^p {\rm d}\lambda}
\end{equation}
with $p=-1$ reflecting the photon-counting weighting that corresponds to the table's per-photon response convention. The standard quadrature rule used to evaluate this integral interpolates onto the bandpass grid, though for template spectra which are sampled more finely than the bandpass (for instance, input spectral templates which contain resolved emission lines), this may lose of order $1\%$ of the in-band flux, so in these instances we choose to employ a cumulative-integral formulation which preserves the cell-averaged flux density. This manifests as two static fixed grids depending on the construction of the input spectra to be rendered, that is, a configuration step that does not affect the differentiability.

The rendering step proceeds as follows. Sources are first projected into detector coordinates and those whose PSF stamps do not overlap the window region are excised. In-band fluxes are evaluated according to Eq.~\ref{eq:lvfflux}, and then each source's PSF is interpolated from a grid of field positions onto a canvas oversampled by a factor of $18\times$ relative to the native $6\farcs15$ pixels. Stamps are deposited at their sub-pixel positions on this oversampled grid using a separable interpolation kernel scaled by the in-band flux, and each $18\times18$ block re-summed to native resolution, conserving flux exactly. Writing the source's sub-pixel position $u$ along each axis as an integer-sub-pixel-cell part $c = \lfloor u \rfloor$ and a fractional part $r = u - c$, the PSF $\mathcal{P}$ is deposited as
\begin{equation}\label{eq:deposit}
  \mathcal{I}(m) = \sum_k w_k(r)\,\mathcal{P}(m - c - k),
\end{equation}
where $m$ indexes oversampled cells within the pixel, and the weights obey $\sum_k w_k = 1$ and $\sum_k k\,w_k = r$, preserving the flux and centroid exactly.

The final ingredient of our rendered scene is a zodiacal light background, evaluated from a reimplementation of the standard interplanetary dust model~(\citealt{kelsall1998}, the same model as is used in the SPHEREx Sky Simulator) and read from a precomputed table. The model is differentiable in sky position, wavelength, epoch and overall amplitude, but not in the parameters of the underlying dust model, as this table is fixed. The zodiacal emission model is evaluated per-pixel in wavelength while holding the approximate field location fixed; this imposes the implicit assumption that the overall level of zodiacal emission does not vary strongly over the $1.1$ arcminutes subtended by a single window, while accounting for the potentially significant spectral variation over the detector $y$-dimension in the window. Zodiacal emission is largely removed by the background term $b$ of Eq.~\ref{eq:linear}, so this is used to define a photon-noise contribution to each measurement's uncertainty, rather than injecting a contaminant. The overall impacts of zodiacal emission on the SPHEREx measurement are important enough to be considered on their own terms in a dedicated future work~(H. Hui et al. 2026b, in preparation).

\eosp\ also features a fast, approximate direct-catalog generation mode that allows the construction of response templates without passing through the full measurement operator (Section~\ref{sec:densmodmaps}). This path calculates the observed noise using an analytic model based on the background and (zodiacal light-dominated) photon noise level, and uses an approximate iterative approach to quickly calculate the effects of blending. Since forced photometry is linear in the source fluxes (a property on which Section~\ref{sec:composition} relies), blending flux contamination is a fixed sparse cross-talk operator $K$ with elements $K_{ij} = g(r_{ij})$ for $i \neq j$, where $g = k \circledast p$ is the matched-filter kernel $k$ convolved with PSF $p$, evaluated at the pair's detector-pixel separation and tabulated. The measured and true fluxes are then related by $\mathbf{f}_{\rm meas} = (\mathsf{I} + K)\mathbf{f}_{\rm true}$. The deblended fluxes follow by the Neumann series ${(\mathsf{I}+K)}^{-1} = \sum_m {(-K)}^m$, truncated at a chosen order and converging to the exact linear forced-photometry solution for the recoverable set. This series converges when the total fractional flux leaked into source $i$ by its neighbors is less than unity ($|K|<1$), which we enforce by requiring a source's row sum $\sum_j K_{ij} < 0.9$, and treating sources which exceed this in isolation. This may be a less-robust approximation for the most dense and blended scenes, which we regard as a limitation of this fast-but-approximate approach. In practice, for sources unaffected by this criterion the series converges satisfactorily ($\lesssim 1 \%$ to truth fluxes in highly blended scenes) within about a dozen iterations on a typical field, with only about a $\sim 5\%$ performance degradation. This approach, though approximate, is fast and differentiable, aligned with the overall intent of the direct-catalog mode which can be used to quickly probe effects at scale or generate synthetic catalog realizations.

Sources that have been designated as extended are rendered by convolving a S\'ersic profile with the local PSF in Fourier space. The key to making this performant is to split what would na\"{\i}vely be a four-dimensional integral for each pixel (over profile, PSF, bandpass, and position on the LVF) into separable components, which is enabled by the instrument design. The shape of the source is a property of the sky, and the instrument PSF one of the optics; the spectral response, on the other hand, is a function of where the photon lands on the detector after it has interacted with the PSF.\ A source centered at $(x_s, y_s)$ then contributes to the native SPHEREx detector pixel at $(x,y)$
\begin{equation}\label{eq:sersicsep}
  I(x,y) = C(x,y)\,\bar{f}_\nu(x_s, y),
\end{equation}
where $C(x,y)$ is the unit-flux S\'ersic profile convolved with the PSF and integrated over the pixel, and $\bar{f}_\nu(x_s,y)$ is the in-band flux density of Eq.~\ref{eq:lvfflux} evaluated through the LVF response at detector position $(x_s, y)$. The response varies along the detector $y$-dimension, that is, the direction of the LVF gradient, while along detector-$x$ we approximate it as being held at the source's position. This neglects the slight curvature or ``smile'' of the LVF channels across detector-$x$ and variations due to optical distortion, but these are small over the extent of a single measurement window.

We assemble the spatial factor $C$ in Fourier space on the $18\times$-oversampled grid as
\begin{equation}\label{eq:sersicft}
\begin{aligned}
  \widehat{C}(\boldsymbol{\kappa}) &=\\ \widehat{\mathcal{P}}(\boldsymbol{\kappa})\;
  &\widehat{S}_n\!\left(r_e\,|\mathsf{A}\boldsymbol{\kappa}|\right)\;
  \mathrm{sinc}\!\left(\frac{\kappa_x\delta}{2\pi}\right)\mathrm{sinc}\!\left(\frac{\kappa_y\delta}{2\pi}\right)\;
  \widehat{W}(\boldsymbol{\kappa};\mathbf{x}_s).
\end{aligned}
\end{equation}
Here, $\boldsymbol{\kappa}$ is the angular wavevector, $\widehat{\mathcal{P}}$ the transform of the PSF stamp interpolated at the source position, $r_e$ the half-light radius, and $\widehat{S}_n(q)$ the radial zeroth-order Hankel transform of the unit-flux S\'ersic profile of index $n$ as a function of dimensionless wavenumber $q = r_e|\mathsf{A}\boldsymbol{\kappa}|$. The matrix $\mathsf{A}$ applies the source's ellipticity, rotated from the sky frame to the detector frame. The sinc functions integrate the light over each oversampled sub-pixel cell of side $\delta = 0\farcs34$, and $\widehat{W}(\boldsymbol{\kappa};\mathbf{x}_s)$ is the exact transform of the same sub-pixel deposition weights $w_k$ of Eq.~\ref{eq:deposit} used in the point source version of the calculation. Each source is rendered on a square canvas centered on it, and a single inverse fast Fourier transform (FFT) gives the flux of the canvas over which the $18\times18$ oversampled cells are summed to produce $C(x,y)$. $\widehat{S}_n$ has no closed form, so we construct a table over $0.3 \le n \le 6.2$ and $10^{-4} \le q \le 10^{3}$ and evaluate $\widehat{S}_n$ using a cubic interpolation on it. This process is then differentiable in $n$, $r_e$, ellipticity, position, and flux, and its cost largely independent of $n$ (computational cost is set only by the canvas size, which may need to be larger to accommodate the most extended profiles). This setup and these approximations are validated against renders produced by \galsim\ for a variety of profiles in Section~\ref{sec:eosphoros:parity}. 

\subsection{Relationship to the SPHEREx Sky Simulator}\label{sec:eosphoros:skysim}

\eosp\ shares significant lineage and much of its instrument description with the SPHEREx Sky Simulator, but the two are somewhat different constructions with markedly different goals. While the Sky Simulator incorporates realistic detector readout, noise sampling, and image-level artifacts to produce a full exposure as would have been seen by a SPHEREx detector, \eosp\ instead produces an idealized rendered scene which would be handed directly to the measurement operator to obtain the response of some input perturbation. They nonetheless share the same tabulated spectral response, default point-spread function inputs, zodiacal background model, and survey plan.

The Sky Simulator contains a ``\texttt{quickcatalog}'' direct catalog simulating module which is the most direct analog of \eosp. Both obviate the image and stamp-generation steps to produce catalog fluxes directly based on a PSF and noise model, but \eosp\ pushes to further extremes in performance by way of additional approximations and by leveraging the performance of GPUs.\ \eosp\ utilizes an analytic, deterministic noise model as described in Section~\ref{sec:eosphoros:rendering}, rather than performing many Monte Carlo draws (500, at the Sky Simulator/\texttt{quickcatalog} default), the precomputed zodiacal emission table, and heavily leverages the GPU acceleration unavailable to the CPU-only \texttt{quickcatalog}. The catalog-level point-source forward model matches that of the Sky Simulator to a relative level of $10^{-12}$ in both flux and wavelength, and the analytic uncertainty reproduces the Sky Simulator's sampled uncertainty to within its sampling floor for the 500 noise realizations. The Sky Simulator does not incorporate automatic differentiation, so the differential response templates enabled by \eosp\ are also a new capability.

The extended source-rendering matches at the percent level, driven by known conventional differences listed below, and the implemented additional optical features (an extended PSF halo and a dichroic ghost, see \citealt{korngut2026}) both match to $10^{-14}$ relative. The three deliberately different conventions are the bandpass flux weighting (evaluated per photon rather than per unit energy), the centering of the PSF stamp, and the sign convention of the optical distortion model. The latter two of these in particular were conventions derived to match flight data, which was unavailable to the Sky Simulator at the time of its creation. The overall differences are smaller than the absolute calibration error (M. Ashby et al. 2026, in press) and of order the relative calibration error.

\subsection{Differentiability}\label{sec:eosphoros:diff}

The \eosp\ renderer is differentiable with respect to source flux, source position, and the systematic operators that it implements. The flux is straightforward and the systematics are situation-specific, but the position in particular requires special care.

We separate the way that the position enters the render into a discrete component, information such as which sources are included in the scene and which window each fell into, which tabulated ePSF position it gathers, which row of the tabulated filter map it reads, and which oversampled cell it lands in, and a continuous component, the sub-pixel offset within a cell. We freeze the discrete component and differentiate the continuous part. This continuum must itself be smooth enough to differentiate, constraining the nature of the kernel. A piecewise-linear deposition with knots at every oversampled cell boundary would have a position derivative piecewise constant and jumping across the knot; on a real ePSF, the derivative of the native peak pixel changes sign across such a boundary. A continuously differentiable kernel removes this feature: within a cell, its position derivative matches a central finite difference at second order, $3.7 \times 10^{-5}$ of the largest derivative at a step size of $10^{-3}$ native pixels, falling to $3.7 \times 10^{-7}$ at a step of $10^{-4}$. Across a cell boundary, it converges at first order, while a piecewise-linear one does not converge at all over an order of magnitude in step sizes.

\subsection{External Validation}\label{sec:eosphoros:parity}

\begin{figure*}
  \centering
  \includegraphics[width=\textwidth]{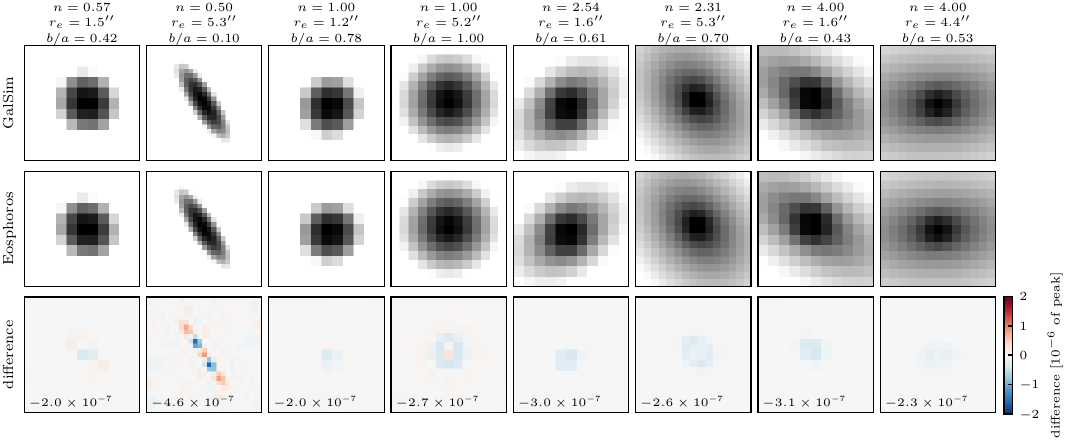}
  \caption{The \eosp\ extended render for eight of the sources from the ensemble of Appendix~\ref{ap:gsval}, over a range of S\'ersic indices and sizes. The top two rows show the render from each code on native SPHEREx image pixels, normalized to the peak with a common logarithmic color scale. The bottom row shows the difference in units of $10^{-6}$ of the \galsim\ peak, with the inset showing the fractional flux difference $\delta f/f$ measured from the two images by the same profile-weighted fit.}\label{fig:validation}
\end{figure*}

A representative sample of eight \galsim~\citep{rowe2015} renders at a range of S\'ersic indices, axis ratios, and position angles is shown in Figure~\ref{fig:validation}. Agreement (maximum deviation in units of the stamp peak) was measured as $1.5 \times 10^{-7}$ to $2.3 \times 10^{-6}$, with fit fluxes typically agreeing to a few parts in $10^{7}$. Detail on the extended profile validation with a larger ensemble of extended profiles is presented in Appendix~\ref{ap:gsval}.

We also validate the kernel itself on its own terms using an identity. The same real tabulated stamp is applied twice to an identical profile representation, once through the Fourier-domain path the renderer uses, and once through an explicit cyclic summation in the image domain. This represents the same operation written two ways, with the two methods agreeing to a few parts in $10^{15}$ relative in their fluxes and uncertainties. Deliberately mis-matching a convention in one or the other dramatically breaks this agreement, as expected.

Though this establishes the profile, kernel, and conventions, it does not evaluate the ePSF model, instrument spectral response, or calibration against that of the real detector. This we will revisit in Section~\ref{sec:corpus}.

\subsection{Throughput}\label{sec:eosphoros:perf}

Rendering a full-sky catalog on a single GH200 with \eosp\ sustains $1.01 \times 10^4$ sources per second in the wide survey regime, using realistic input spectra from \citet{feder2024}. A single observation takes a few tenths of a microsecond on average to process on the GPU, with a peak of $18.5$ GB device memory. This was found to be linear with catalog size over two orders of magnitude, with a $10^8$-source pass costing approximately $2.7$ GH200 node-hours. An end-to-end run through the full differentiable measurement described in Section~\ref{sec:densmodmaps}, consisting of rendering, rebinning, redshift estimation, tomographic assignment, and the Jacobian-vector product on $10^8$ sources took $5.78$ node-hours, making the full pipeline about $2.1\times$ as time-consuming as the catalog forward mode alone. On identical CPU hardware, the Sky Simulator \texttt{quickcatalog} rendering at default configuration was measured at $88.6$ ms/observation, compared to 1.78 $\mu$s for \eosp, a factor of $\sim 5 \times 10^4$. We note that a large component of this is the Monte Carlo noise draws performed by the Sky Simulator (500 at the default configuration against which we compare), which are a tunable parameter, as compared to the analytic noise of \eosp. These measurements are tabulated in Table~\ref{tab:eosp_throughput}.

\begin{deluxetable*}{llrrr}
\tablecaption{\eosp\ catalog throughput compared with the Sky Simulator\label{tab:eosp_throughput}}
\tablehead{
  \colhead{Code} & \colhead{Hardware} & \colhead{Time per observation} &
  \colhead{vs.\ \texttt{quickcatalog}} & \colhead{$10^8$ sources (hours)}
}
\startdata
\cutinhead{Catalog forward model}
\texttt{quickcatalog}, 500 noise draws & 16-core CPU & 88.6 ms & 1 & $1.2\times10^6$ \\
\texttt{quickcatalog}, 100 noise draws & 16-core CPU & 19.8 ms & 4.5 & $2.7\times10^5$ \\
\eosp & 16-core CPU & 1.78 $\mu$s & $5.0\times10^4$ & 24 \\
\eosp & $1\times$ GH200 & 0.20 $\mu$s & $4.4\times10^5$ & 2.7 \\
\cutinhead{Full differentiable measurement pipeline}
\eosp & $1\times$ GH200 & \nodata & \nodata & 5.8 \\
\enddata
\tablecomments{Wide-field sources, about 490 observations each over the survey. CPU rows are the same 16-core Apple M4 Max laptop CPU~\citep{m4max}. The \texttt{quickcatalog} $10^8$-source times are extrapolated from its per-observation rate, which is linear in the number of noise draws. The full differentiable measurement pipeline, described in Section~\ref{sec:densmodmaps}, consists of source rendering, flux binning, redshift estimation, tomographic assignment and the Jacobian-vector product.}
\end{deluxetable*}

The forward model and the measurement (Section~\ref{sec:hesperos:perf}) are affordable on the same scale, naturally facilitating their composition. The ability to forward model the measurement, and specifically the differential response to systematics, at this scale changes what systematics templates can be, as well: rather than regressing against a proxy of the effect which might couple to the density field, we may directly measure its response, limited only by the model of how the systematic couples to the measurement itself. For systematics whose effect on the measurement is easily modeled (\emph{e.g.} observing depth as dictated by zodiacal emission), this largely alleviates the potential bias that systematic template deprojection cleaning approaches can incur due to decorrelation between the physical systematic's coupling to the density field and the regressor proxy. This aspect is elaborated upon through a composition with the redshift estimation and sample selection demonstrated in Section~\ref{sec:densmodmaps}.

\section{Composing the model and the measurement}\label{sec:composition}

The forward model and the measurement operator are composed by JIT-fusing a \eosp\ scene render to a \hesp\ measurement and differentiating the result. This section describes what is necessary to enable this, what the derivative represents, and the conditions under which it is valid.

\subsection{Measurement geometry}\label{sec:composition:geometry}

The Level 2 image source catalog containing the SPHEREx Reference Catalog sources to be photometered, or the injection catalog, is provided directly to the measurement engine \hesp\ by virtue of the forced-photometry measurement. We use these same positions and the same L2-provided tangent-plane-with-distortion parameters to render the sources with \eosp, ensuring co-location of the rendered sources to their measured positions. A fallback mode utilizes the pre-flight tabulated survey plan rather than as-flown pointings and positions. A coherent offset of one eighteenth of a native pixel ($0.34''$) produces a few tenths of a percent in flux error, while a full pixel offset reduces the measured flux for a force-photometered source by about 15\% on average. The measurement geometry of \hesp\ and \eosp\ are verified to produce identical results, with WCS/optical distortion, sub-pixel placement, and position derivatives verified at or better than $1\times 10^{-8}$ pixels.

\subsection{Differentiable significance threshold}\label{sec:composition:psig}

We define as a figure of merit the probability that a measurement is significant, that is, that its fitted flux $\hat{f}$ exceeds a threshold $\nu \sigma^{\rm marg}$,
\begin{equation}\label{eq:psig}
  P_{\rm sig} = \Phi \left( \frac{\hat{f}}{\sigma^{\rm marg}} - \nu \right),
\end{equation}
with $\Phi$ the standard normal cumulative distribution function and $\hat{f}$ the noise-free expectation of the flux measurement. This assumes a Gaussian distribution of flux errors, an approximation which may be strained in the presence of un-flagged image artifacts, but is sufficient for this image-level demonstration. The naturally imposed smoothing scale $\sigma^{\rm marg}$ helpfully makes an otherwise discontinuous thresholding problem differentiable. We adopt $\nu = 5$, and use the full marginal uncertainty to account for the appropriate significance for blended sources. Note that this is a per-exposure, per-measurement figure, not necessarily a reflection of whether a given source is ``detected'' insofar as it is measured with sufficient precision to enter the redshift sample, which is a function of all of the measurements of that source in all the bands in which it is sampled. The quantity reported in Section~\ref{sec:corpus} is correspondingly $\mathrm{d}\ln N_{\rm sig}/\mathrm{d}\alpha$, a response of the image-level spectrophotometric measurements rather than of the cosmological sample.

\subsection{Differentiation}\label{sec:composition:diff}

A systematic amplitude $\alpha$ may elicit a response in $P_{\rm sig}$ through its flux and its uncertainty, and which of those routes (or both) that it enters is mechanically relevant for the automatic differentiation. An $\alpha$ that modifies the flux of a scene changes the input data $\mathbf{d}$ without changing the covariance (pixel weights) or the ePSF model used by the fit. The entire derivative then flows through $\hat{f}$, which we refer to as the ``flux path''. If $\mathbf{d}$ changes linearly with $\alpha$, such as for a fixed zero-point offset or additive background, the response is also linear and the derivative is exact at any amplitude. If $\mathbf{d}$ instead depends nonlinearly on $\alpha$, then the response inherits that order; for a perturbation in detector position, the response is the derivative of the overlap of the rendered profile and the profile assumed by the fit, which is maximized at zero offset and vanishes at first order for a continuous profile, the flux loss being second order (though on a pixel grid, a residual first-order term set by the sub-pixel phase may still enter, see Section~\ref{sec:corpus}). An $\alpha$ that affects the noise, rather than directly changing the flux of an image, such as the zodiacal background level, acts through the inverse-variance weights $W$ and therefore through $\mathrm{N}$; we refer to this as the ``variance path''. A systematic that affected the fluxes, such as Galactic extinction, would correspondingly operate on $\mathbf{d}$ directly through the ``flux path''. A change in the inverse-variance weights changes the flux by $\delta\hat{\boldsymbol{\theta}} = \mathrm{N}^{-1}\mathrm{A}^{\rm T}\,\delta\mathrm{W}\,(\mathbf{d}-\mathrm{A}\hat{\boldsymbol{\theta}})$, that is, proportionally to the difference of the data and the best-fit model (a window whose pixels disagree on its best-fit solution is the most prone to having that solution perturbed by a change in the relative weights of those pixels). A uniform change in the weights $\mathrm{W}\to(1+\epsilon)\mathrm{W}$ would produce no change in the flux, but would scale every marginal uncertainty as $\sigma^{\rm marg}\to\sigma^{\rm marg}/\sqrt{1+\epsilon}$, such that
\begin{equation}
  P_{\rm sig} = \Phi\left(s\sqrt{1+\epsilon}-\nu\right), \qquad s \equiv \hat{f}/\sigma^{\rm marg} \text{ at } \alpha = 0,
\end{equation}
with $s$ the source's measured significance. A non-uniform change in the weights, such as a variation in the zodiacal background across a window, could change the fluxes; the automatic derivative captures this response in any case. However, this derivative is an adequate description of the response only when the shift in $s$ is small compared to the width of $\Phi$ near the threshold; if every pixel's variance were to be changed by the same factor, this would require $\epsilon \ll 2/\nu$.

A positional perturbation will operate via the flux path, since the fit's source profiles are fixed at the catalog positions for forced photometry and only the rendered image is affected. Uniquely for SPHEREx, though, this is not solely geometric: the linear variable filter responds in wavelength along the detector-$y$ direction, tying bandpass to detector position. A displacement along the dispersion axis of the LVF then moves its flux relative to the fit's profile and affects the wavelength at which that flux is emitted into the scene. Astrometric and spectral responses are coupled via the LVF, and both must be taken into account for both the overall response and the derivatives. In the case of a matched model for a continuous profile, this spectral term is the only one that contributes at first order, the geometric flux loss being second order. However, a first-order geometric term survives on a real pixel grid, and is the dominant term in the response in five of the six arrays (Section~\ref{sec:corpus:response}). The spectral term is
\begin{equation}\label{eq:lambdaderiv}
  \left.\frac{\mathrm{d} \ln \hat{f}}{\mathrm{d}y}\right|_{\rm spec} = \frac{\mathrm{d} \ln \bar{f}}{\mathrm{d}\lambda} \frac{\mathrm{d}\lambda}{\mathrm{d}y}
\end{equation}
for a shift along the LVF gradient direction $y$. As we see from the first term on the right-hand side, the steepness of the spectrum being measured also plays a role in setting the amplitude of this spectral contribution to the gradient, as does the LVF response as represented by the second term. The coupling of position and wavelength dependence is also relevant for slitless spectroscopy, typically treated as a calibration term (\emph{e.g.} \citealt{gillard2026}).

With a single scalar input systematic amplitude and per-measurement probabilities $P_{\rm sig}$ as output, we would like the directional derivative of a high-dimensional output with respect to a low-dimensional input, so forward-mode differentiation is the natural choice. The dominant cost is a one-time initial step, that of building the initial linear map, after which it may be applied to multiple directions cheaply. A single derivative direction costs around $2.4\times$ a plain render, in addition to the $\sim 6\times$ needed to build the linear map once. Overall, for an ensemble of sources, a single direction adds about $29\%$ to the runtime versus a typical forward pass of the render and measurement, with additional directions adding $\sim 8\%$ each.

Note that the calculated derivatives are valid within only the regime of the frozen aspects of their configuration, such as which sources exist, which window each is central in, the bucket and chunk each belong to, assigned ePSF zone (a detector sub-region over which a single ePSF model is defined), and so on. Practically, aspects such as the window membership and source list are fixed by the catalog, and bucket/chunk assignment will only affect floating-point ordering; tabulated quantities based on rendered position, such as the ePSF zone, will be crossed under a coherent shift $\delta$ by a fraction $\approx\delta/L$ of sources, where $L$ is the zone spacing, $\approx 100 $ native pixels. For a coherent shift of $0.1$ detector pixels, this corresponds to about $0.1\%$ of sources entering a different zone. A perturbation large enough to change a source between one of these states will not be captured by the automatic derivative, and in our implementation raises an error.

\subsection{Verification}\label{sec:composition:val}

Because our fit is linear in the image, we may construct an exact test of this composition: adding source $i$'s own unit-flux model to the image should change the fitted flux of $i$ by exactly one unit, and the fitted flux of every other source in the window by exactly zero. Rendering source $i$ with \eosp\ and differentiating the \hesp\ fit along that direction passes this test to a level of $10^{-9}$ relative for an isolated source, a blended pair sharing a window, and a pair in distinct overlapping windows. We test the reverse-mode gradient similarly, and the gradient of the fitted pixels with respect to the input functional form matches to one part in $10^8$, including blended windows and the background. This verifies both the mechanics and the content of the operator along the desired direction, but we additionally verify the former properties against a central finite-difference (FD) approach. For the flux path, in particular, the FD can be shown to be exact with no step-size error by virtue of the fitted flux response to a scene amplitude being linear. For this case (and for the variance path for which we cannot establish such an identity), the AD and FD derivatives are found to be in agreement to less than one part in $10^9$.

The spectral term of Equation~\ref{eq:lambdaderiv} is also validated with both an identity and an empirical central finite-difference check. For a source with a power-law SED, $f \sim \lambda^\beta$, $\mathrm{d} \ln f / \mathrm{d}y = \beta \cdot \mathrm{d} \ln \lambda_0 / \mathrm{d}y$, which we may immediately relate to Equation~\ref{eq:lambdaderiv}. Three toy power-law input spectra were used, with $\beta$ of $-1.2$, $0.4$, and $-2.0$. For all of these, the agreement of the measured derivative to the identity is 0.6\%, with the residual disagreement being attributable to the interpolation of the SED off of a finite grid (the identity applies exactly only for a continuous integral). The FD validation matches the AD Jacobian-vector-product to 2 parts in $10^9$ relative at a step size of $10^{-3}$ native pixels. The FD validates only the wiring, not the physics, so again both are needed.

These verify that the derivative propagates correctly through the operator, though do not validate those operators themselves, which is the subject of Sections~\ref{sec:hesperos:valtractor} and~\ref{sec:eosphoros:parity}.

\section{From fluxes to a three-dimensional density field}\label{sec:densmodmaps}

Photometry is a single step of the measurement pipeline. The ultimate product, from the point of view of a cosmologist, is not a catalog of fluxes but a representation of the three-dimensional galaxy density field. Accordingly, the response we are interested in is that of a perturbation on that density field through the full pipeline. This section describes the remaining measurement steps.

We first note that there are two possible routes to this downstream density field: the composition of the renderer with the full measurement operator as presented in Section~\ref{sec:composition}, and the fast, approximate, direct-catalog (\eosp-only) model described in Section~\ref{sec:eosphoros:rendering}. The former functions at a per-pixel level based on actual as-flown frames, using the real measured background and variance layer; the latter uses fast approximations for both the survey plan/pointings and the measurement noise. This is easily scaled to high densities and able to be used to predict the response over the full survey (including observations which have not yet happened), so is suitable for the generation of the response templates described in this section. For the depth systematic presented in this section, the two were compared on a common scene with common exposures, where the fit is exact and so the flux is recovered identically, isolating the analytic noise model. Across $422{,}492$ measurements over 125 exposures and all six detector arrays, the direct-catalog per-measurement uncertainty has a median ratio of 0.997 to the image-level uncertainty, with per-array medians within 1\% and a 16th--84th percentile spread of 0.71--1.15. The depth response is a closed-form function of the flux and this uncertainty, so its agreement follows. This is the desired behavior, with the full composition of Section~\ref{sec:composition} being necessary to characterize effects which have detailed per-pixel manifestations, or to probe regimes where the analytic models break down; this is discussed further in Section~\ref{sec:discussion:limitations}. The responses described in this section use the fast, direct-to-catalog mode.

\subsection{Fiducial photometry channels}\label{sec:densmodmaps:fiducial}

The positions at which flux measurements land on the LVF are continuous, therefore landing at a continuous distribution of wavelengths, so-called ``native'' flux measurements. An initial step of the pipeline is to weight and bin these into the 102 ``fiducial'' channels on which the fixed-grid redshift fitting is performed. Higher fiducial bin counts corresponding to higher spectral resolution, for instance, 306 bins, are used for some applications.

Each measurement is assigned to the fiducial filter with the nearest central wavelength with a weight according to the observing wavelength and that filter's transmission curve $T_{\rm norm}$, normalized to peak. Fluxes are combined with standard inverse-variance weighting $T^2_{\rm norm} / \sigma^2$. Unobserved channels are assigned zero weight, which accounts for uneven spectral coverage. The fiducial bin assignment is itself a static geometric feature, and derivatives pass only through the weighted co-add.

The reimplementation of the channel binning is validated at a relative tolerance of $10^{-10}$ on flux and uncertainty against the existing L4 reference implementation, with the list of surviving measurements (pattern of dropped/flagged/NaN measurements) held to match exactly.

\subsection{Redshift estimation}\label{sec:densmodmaps:photoz}

The redshift estimation is based on a standard template-fitting approach, reviewed here at a necessary surface level. The spectrum of binned photometric measurements is compared to a list of spectral templates shifted to a variety of redshifts. For a source $i$ and a template $m$ redshifted to grid node $z$, the amplitude is profiled analytically, giving
\begin{equation}\label{eq:chi2}
  \chi^2_{im} = C_i - B^2_{im} / D_{im}
\end{equation}
with $C_i = \sum_f F^{f\ 2}_i w^f_i,\ B_{im} = \sum_f F_{i}^f f^f_{m} w^f_{i}$ and $D_{im} = \sum_f (f_m^f)^2 w_i^f$ over fiducial channels $f$ with weights $w = 1/\sigma^2$. The redshift probability density is then $P(z) \propto \sum_m P_{\rm tmpl}(m) \exp(-\chi^2_{im}/2)$, where $P_{\rm tmpl}$ is the prior weight for spectral template $m$ (for SPHEREx, these priors are taken to be uniform). Both $B$ and $D$ are batched matrix products, so the fit over the catalog is two large contractions rather than a loop, enabling significant acceleration on GPU for similar reasons as described in Section~\ref{sec:hesperos:perf}.

This template-fitting estimator is in the tradition of the code \textsc{LePHARE}~\citep{arnouts1999,ilbert2006}, with a SPHEREx implementation descending from a variant initially described in \citet{stickley2016} and updated in \citet{feder2024}. The production SPHEREx redshift pipeline, its development status, and performance will be the subject of a future publication (D. Masters, Z. Huai et al. 2026, in preparation). As a spectral template basis, we use the library of \citet{feder2024}. These are the same spectra as are used to render these sources per Section~\ref{sec:eosphoros:rendering}, ensuring the redshift fitting is self-consistent\footnote{For actual cosmological measurements and templates that would be used in the final analysis, we plan to use internally-constructed empirical galaxy SED templates, which will themselves be the subject of a forthcoming work~(Z. Huai et al. 2026b, in preparation).}.

We validate the redshift fitting implementation directly against the Level 4 production C++ implementation, with the two agreeing to $1.5 \times 10^{-5}$ relative in the redshift posterior $P(z)$ maximum deviation, within the reference implementation's own (32-bit floating point) precision.

\subsection{Tomographic assignment}\label{sec:densmodmaps:tomo}

A potentially non-differentiable step is the estimation of redshift and assignment to a tomographic bin. The redshift estimate is typically taken as the \emph{maximum a posteriori} or, in the current implementation's case, the expectation value of the redshift probability density function (PDF). Even if an expectation value calculation may itself be differentiable, the bin assignment necessary for practical use may still prevent practical differentiability. Instead, we assign a source $i$ with a vector of flux measurements $\mathbf{F}_i$ to a tomographic redshift bin $b$ according to the mass of its posterior falling in each bin,
\begin{equation}\label{eq:tomo}
  s_{ib} = \frac{\int_{z_b}^{z_{b + 1}}P(z|\mathbf{F}_i ) \mathrm{d}z}{\int P(z | \mathbf{F}_i ) \mathrm{d}z}.
\end{equation}
The bin edges only enter through a static mask, so the derivative flows through the normalized posterior and the assignment remains differentiable. Posterior mass outside of the considered bins is unselected, so a source's memberships sum to at most unity rather than being renormalized onto the redshift grid.

\subsection{Density field response}\label{sec:densmodmaps:maps}

\begin{figure*}
  \centering
  \includegraphics[width=\textwidth]{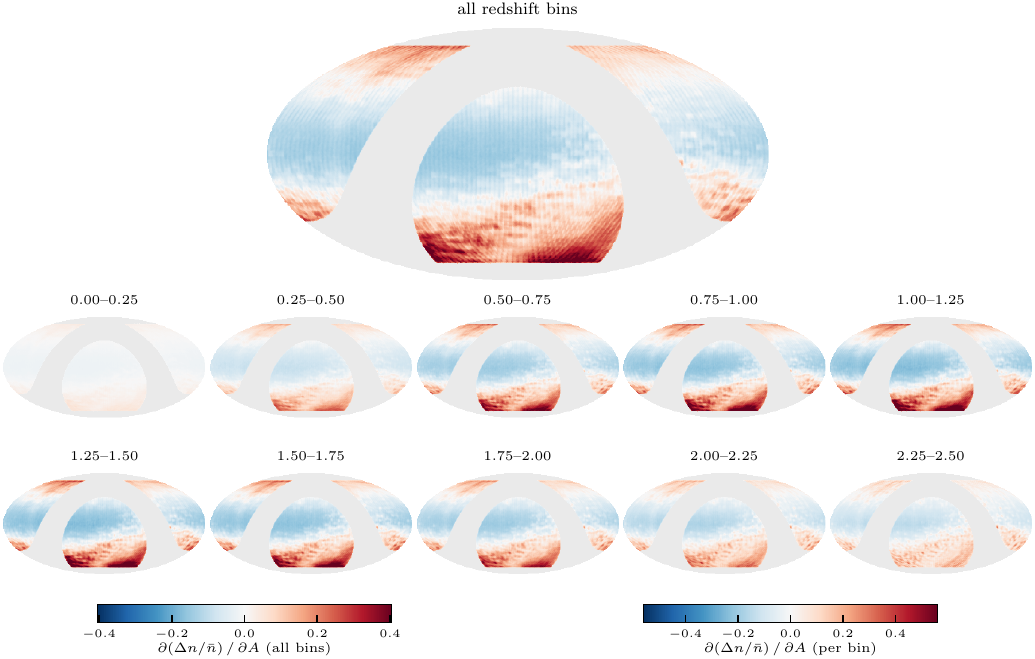}
  \caption{The density field response to unit variation in the survey-depth systematic $\partial(\Delta n / \bar{n})/\partial A$ for the SPHEREx wide-field survey using a simulated 2-year survey plan, forward-modeled with the direct-catalog mode of \eosp. $A=0$ corresponds to uniform coverage, and $A=1$ to the realized depth as described in Eq.~\ref{eq:ampmod}; the template shows the value of the derivative evaluated at the realized depth, the output product of the full composed analysis pipeline. The top row shows an aggregate over radial bins and is smoothed at $1^\circ$ for visualization while the middle and bottom rows show individual $\Delta z = 0.25$ bins from $0$ to $2.5$ and are smoothed at $2^\circ$. Both are zero-mean over the sky. Maps are shown in ecliptic projection and a $|b| < 20^\circ$ Galactic latitude cut is imposed, and a cut is also placed on the $\sim 50\times$ deeper SPHEREx survey ``deep fields'' in the ecliptic poles beyond ecliptic latitude $|\beta| > 70^\circ$, to showcase the primary area for the cosmology analysis and to avoid dominating the color scale. Sources are drawn uniformly over the sky; the visual structure is a convolution of
    the survey hit pattern and wavelength-/detector-position-on-FoV-specific optical effects with the zodiacal emission model which sets the photon noise floor in a given exposure. 0.1\% of voxels whose gradients left the linear regime have been excluded. Note that this represents the effect on a uniform distribution of sources in comoving volume, so may not be representative of the response of an actual SPHEREx sample selection. This template is built from $10^{8}$ total sources, or $\sim 6 \times 10^{10}$ total measurements, run in about 6 node-hours via the fast catalog-direct (\eosp-only) mode on a GH200.}\label{fig:template}
\end{figure*}

Each measurement of a source has a probability $P_{\rm sig}$ of being significant (Equation~\ref{eq:psig}), and an ensemble of measurements across a spectrum are assigned to tomographic bins according to Equation~\ref{eq:tomo}. Final redshift fitting-based selections are applied at this stage, treated statically (for the current analysis), and then the population of each voxel (pixel-redshift bin combination) is reduced to a fractional density modulation $\Delta n / \bar{n}$, in the same form an angular clustering systematic template analysis would ingest. For SPHEREx, these templates also contain the full spectrophotometric response of the measurement through the measurement pipeline, incorporating radial effects into the response maps directly.

With each stage from pixel data to the three-dimensional density map implemented as a composition of matrix operations on matrices with fixed shapes, JAX allows us to take a single directional derivative through the pipeline and obtain the exact response of the pipeline to that perturbation without need for Monte Carlo sampling, at a cost slightly higher than a single forward evaluation. Figure~\ref{fig:template} shows such a 3-D response measurement for the survey observing depth, represented via a parameter $A$. $A$ is a dimensionless amplitude that modulates the co-added variance for each source in each channel continuously between the population mean $\bar{v}_c$ and its realized value,
\begin{equation}\label{eq:ampmod}
  v_{ic} (A) = \bar{v}_c + A\,(v^{\rm real}_{ic}-\bar{v}_c),
\end{equation}
where $v^{\rm real}_{ic}$ is the co-added variance from source $i$'s actual observations in channel $c$. $A=1$ corresponds to the realized survey, while $A=0$ gives each channel its sky mean variance. The template $\partial(\Delta n/\bar{n})/\partial A$ at $A=1$ is therefore the linear response of the density field to the realized depth variation.

\begin{figure}
  \centering
  \includegraphics[width=\linewidth]{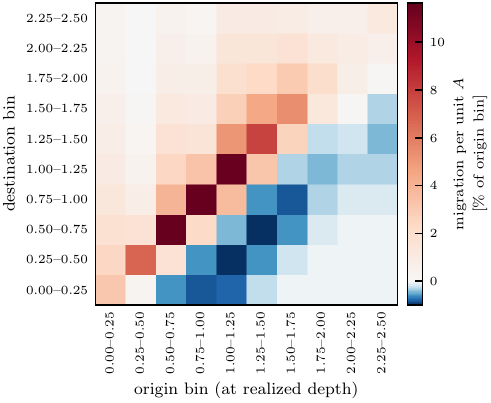}
  \caption{Redshift bin migration matrix for the $\partial(\Delta n / \bar{n})/\partial A$ template, showing the fractional migration per redshift bin per unit $A$, normalized by the origin bin's count. The bright diagonal represents the change in how many of a bin's own sources are selected ($P_{\rm sig}$, and the peak of its smooth posterior from Eq.~\ref{eq:tomo}), while the off-diagonal represents the change in the posterior mass that lies outside of the peak bin, which the modulation shifts from lower to higher redshifts. Note that sources are entitled to leave the sample altogether, so the total number of photometered sources represented within this figure is not necessarily preserved. This plot aggregates physically different noise regimes, and is intended solely to illustrate the off-diagonal coupling induced by the depth variation through the redshift fit.}\label{fig:migration}
\end{figure}

The template is constructed using the analytic noise model and catalog-direct mode of Section~\ref{sec:eosphoros:rendering} which couples the instrument to a prediction of the mean zodiacal emission level (which determines the photon noise budget), convolved with the survey scanning strategy, represented here through a pre-flight tabulated proxy survey plan. These maps are direct measurements of the response of the downstream field to the systematic $\partial(\Delta n / \bar{n})/\partial A$ through the pipeline, rather than a potentially decorrelated proxy of it (subject to the confidence of the input systematic model and fidelity of the calibration products/survey plan). For instance, because it is constructed on a per-observation basis, the significant time variability of the zodiacal emission is straightforward to fully incorporate. Figure~\ref{fig:migration} shows the redshift bin migration matrix resulting from this unit depth variation, illustrating the off-diagonal coupling induced by the spatially non-uniform redshift recovery. The derivative through this full chain of operations, including redshift fitting and tomographic assignment, is verified against a central finite difference on a reduced configuration, with a $6 \times 10^{-4}$ relative error at an FD step size of $0.01$ in $A$, with that residual falling as the square of the step size, as expected.

\section{Survey-scale selection response}\label{sec:corpus}

\begin{figure*}
  \centering
  \includegraphics[width=\linewidth]{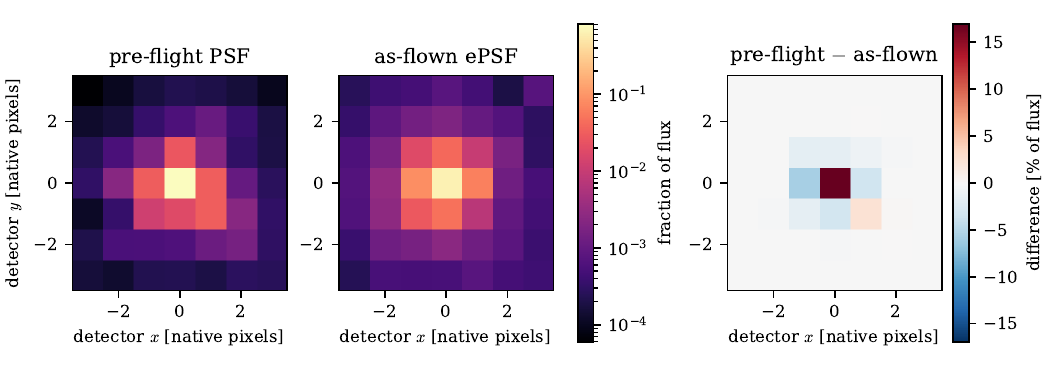}
  \caption{Comparison of the pre-flight (left) and as-flown measured ePSF (center) models, normalized to unit total flux on a logarithmic color scale, with their difference shown in the right-hand panel. In general, the PSF depends highly on focal plane position; the example shown is that centered at pixel (877, 877) of Detector 1, which sits upon one of the tabulated pre-flight PSF values (rather than a point that was interpolated between tabulated values).}\label{fig:psf-comp}
\end{figure*}

This section demonstrates the use of the full \eosp+\hesp\ composition on synthetic scenes using real frame metadata at survey scale, and the type of effect that may be measured. We ran a set of test observations where sources were rendered with the pre-flight PSF model, ray-traced from the as-built optical design~\citep{korngut2026}, but measured with the real effective PSF estimated from reference objects on flight data. In the control, the same flight data-derived ePSF was used for both the render and the solve. These PSF models and their difference are shown in Figure~\ref{fig:psf-comp}. Note that the ePSF model used herein is itself preliminary and subject to change; this is not intended to represent a statement about SPHEREx's actual optical response or PSF model, but rather about this pipeline's sensitivity to a (in this case, known to be) mis-specified model.

\subsection{Setup}\label{sec:corpus:setup}

Synthetic sources are rendered in frames corresponding to $20\times$ $\sim3.5$-day ``planning periods'' spread evenly over the first $\sim 15$ months of data, 216,437 exposures in total across the six detectors. This represents 16.6\% of the full dataset for that time period. Sources were rendered at real Reference Catalog positions for each image, measured with the production fit, and the derivative calculated through $P_{\rm sig}$. Each exposure had $\sim 2,500$ sources, over $\sim 1.1 \times 10^9$ total between both PSF models, completing in $7.40$ and $22.3$ $4\times$GB200 node-hours for the pre-flight and as-flown PSF models, respectively, the difference being due to the larger support and therefore number of pixels required by the ePSF during the render step.

The systematic amplitude $\alpha$ is a coherent astrometric offset in detector pixel units, applied to the scene with the fit's source profiles fixed at catalog positions as described in Section~\ref{sec:composition:diff}. Three different sets of gradients are calculated for each run: coherently shifting every source in a frame in detector-$x$, shifting in detector-$y$ (the LVF gradient direction), and an ensemble of random directions. For a response that is local and linear in the source's own displacement, the random-direction gradient should average to zero over a reasonable ensemble of sources, making it serve as a null test.

Sources are assigned spectra and are resampled to populate redshift bins uniformly, ensuring uniform shot noise. Source amplitudes are calibrated to deliberately straddle the signal-to-noise threshold to better inform the behavior in this more interesting regime. Because the rendered and fitted PSF profiles differ, we do not recover the input signal-to-noise distribution exactly. The difference is additive in signal-to-noise, and for the pre-flight render is a manifestation of the model misspecification under study; we therefore report each source's achieved signal-to-noise, and the response as a function of it.

After all selections, the count in a bin is the expected number of significant measurements $N_{\rm sig} = \sum_i P_{{\rm sig},i}$, so we report
\begin{equation}\label{eq:dlnndet}
  \frac{\mathrm{d} \ln N_{\rm sig}}{\mathrm{d}\alpha} = \frac{\sum_i \mathrm{d}P_{{\rm sig},i}/\mathrm{d}\alpha}{\sum_i P_{{\rm sig},i}},
\end{equation}
in units of inverse native detector pixels. This is a ratio of sums, rather than an average of per-source ratios, which is an important distinction for a population clustered near the significance threshold.

Responses to a coherent perturbation within an exposure are correlated by their frame and blend configuration, so the uncertainty is instead estimated across exposures, with Equation~\ref{eq:dlnndet} formed independently per exposure. The signal-to-noise distribution here is a construction of the experiment rather than a model of any realistic population, which is inherited by the normalization of Equation~\ref{eq:dlnndet}.

\subsection{Response}\label{sec:corpus:response}

\begin{figure}
  \centering
  \includegraphics[width=\linewidth]{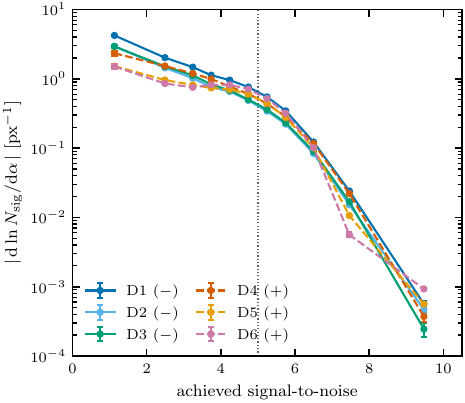}
  \caption{Response of the number of significant measurements to a perturbation along the detector-$y$ (LVF gradient dimension) for each detector when rendered with the pre-flight PSF, as a function of achieved signal-to-noise. The averaged sign of the response is noted in the legend. The dotted line represents the nominal threshold $\nu = 5$. Error bars show the standard error on the mean formed across exposures and are independent. Lower signal-to-noise sources closer to the threshold are more likely to be pushed above it or below it by a given perturbation, while bright sources are significant regardless and show little response.}\label{fig:response}
\end{figure}

Figure~\ref{fig:response} shows the pre-flight-render response as a function of achieved signal-to-noise for each detector array. The response falls by four orders of magnitude over a factor of 10 on achieved signal-to-noise. Low-significance sources near the threshold are the most likely to cross the threshold due to a given perturbation, so this is as expected.

\begin{figure*}
  \centering
  \includegraphics[width=0.8\textwidth]{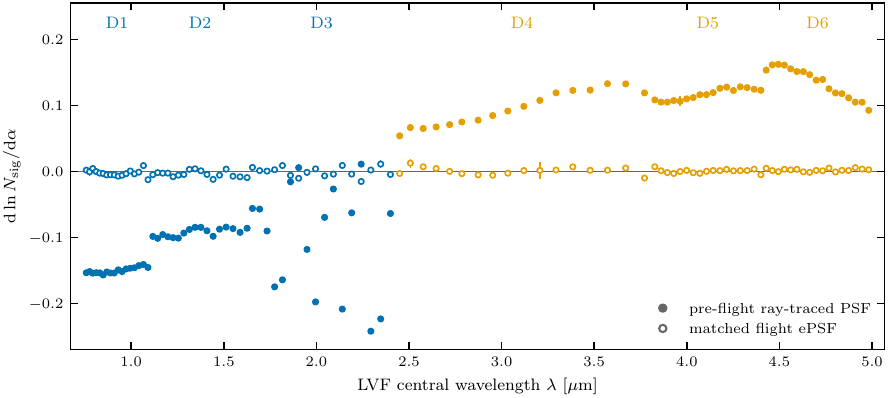}
  \caption{The coherent response $\mathrm{d} \ln N_{\rm sig} / \mathrm{d} \alpha$ along the LVF gradient direction (detector-$y$) for the pre-flight PSF model and the as-flown data-derived ePSF, over the 102 fiducial wavelength channels and the six physical detector arrays. Points represent binned native-wavelength observations with error bars again showing the standard error on the mean across independent exposures. The residual offset in the matched ePSF case is small and is reflective of the LVF response to this ensemble of input SEDs. The significance is against exposure-to-exposure scatter and increases with the square root of the number of exposures used.}\label{fig:response_lambda}
\end{figure*}

Figure~\ref{fig:response_lambda} shows and Table~\ref{tab:response} tabulates the coherent response along the LVF gradient direction for the mismatched-PSF and matched-PSF runs. When sources are rendered with the pre-flight PSF, a coherent shift along the LVF spectral dimension changes the expected number of significant measurements by 9--15\% per native pixel. This is negative on the short-wavelength infrared (SWIR) arrays and positive on the mid-wavelength infrared (MWIR) arrays. Rendering with the ePSF used in the fit reduces this response by a factor of 26 or more on every array, to 0.4\% per pixel or less. Uncertainties shown in Table~\ref{tab:response} are standard errors over $\sim 36{,}000$ exposures per array, and the per-exposure scatter is comparable to the mean response, so the significance is a statement of how well the sample resolves that mean. With the pre-flight PSF, the detector-$x$ response on D1, D5 and D6 is $+5.7$ to $+7.1 \times 10^{-2}$ px$^{-1}$, far beyond the control, and since the LVF model itself has no $x$-dependence over this extent, this is a statement about PSF-shape mismatch on these three arrays.

The detector-$x$ response in the matched case originates from pixel sampling. The first order cancellation of Section~\ref{sec:composition:diff} holds exactly for a continuous profile but is violated by finite pixelizations; on a discrete pixel grid, a sum-normalized profile is peakier near pixel centers, so the first-order derivative of the overlap does not vanish. This term is flat in wavelength within each array, negative on all six arrays, and averages to a small net offset. We do not predict its population amplitude. The additional scatter in the Detector 3 pre-flight response was due to a known optical feature causing the response to vary quickly and non-monotonically with position, which was only properly corrected in the subsequent ePSF model.

For the matched-model control, the detector-$y$ response is resolved on D1, D2, and D6, and consistent with zero on D3--D5. Where resolved, this represents a manifestation of the spectral term of Eq.~\ref{eq:lambdaderiv}, negative where the mean $\mathrm{d} \ln \bar{f} / \mathrm{d} \lambda$ against $\mathrm{d}\lambda / \mathrm{d}y$ is negative and positive where that slope is positive. For the realistic galaxy SED ensemble in use, this means negative responses at short wavelengths where the SWIR observes and positive responses where the MWIR observes, where most galaxy SEDs have turned over.

At this level, the agreement with the matched ePSF is a statement on internal consistency of the pipeline and instrument model, as we have not explicitly described any connection to or grounding in on-sky measurements. This connection has entered implicitly, via the measured ePSF, which was itself created as a direct result of the pre-flight model being found to be inadequate on calibration data based on disagreement with external photometric measurements~(D. Masters et al. 2026, in preparation). We may interpret the inconsistency as requiring the render-side PSF model to be updated, rather than the reverse. The mismatch between the ray-traced and as-flown PSF had been discovered in early photometric validation, but the updated model had not yet been propagated to the Sky Simulator upon which \eosp's renderer was based.

\begin{deluxetable*}{lrccc}
\tablecaption{Coherent astrometric response per detector array\label{tab:response}}
\tablehead{
  \colhead{Detector} & \colhead{$N_{\rm exp}$} &
  \colhead{Coherent $y$} & \colhead{Coherent $x$} & \colhead{Random directions}
}
\startdata
\cutinhead{Pre-flight ray-traced PSF render}
D1 & $36{,}131$ & $-14.98 \pm 0.05$ ($-300.3$) & $+5.70 \pm 0.04$ ($+133.4$) & $-0.02 \pm 0.04$ ($-0.4$) \\
D2 & $36{,}126$ & $-9.21 \pm 0.05$ ($-203.3$) & $-0.22 \pm 0.04$ ($-5.6$) & $-0.04 \pm 0.04$ ($-1.0$) \\
D3 & $36{,}115$ & $-10.25 \pm 0.05$ ($-194.7$) & $-0.32 \pm 0.04$ ($-7.6$) & $-0.03 \pm 0.04$ ($-0.8$) \\
D4 & $36{,}021$ & $+9.53 \pm 0.04$ ($+220.3$) & $-0.38 \pm 0.04$ ($-10.3$) & $+0.01 \pm 0.03$ ($+0.3$) \\
D5 & $36{,}033$ & $+11.70 \pm 0.05$ ($+232.5$) & $+6.84 \pm 0.05$ ($+148.4$) & $+0.02 \pm 0.04$ ($+0.6$) \\
D6 & $36{,}011$ & $+13.53 \pm 0.05$ ($+253.4$) & $+7.14 \pm 0.05$ ($+155.0$) & $-0.02 \pm 0.04$ ($-0.4$) \\
\cutinhead{Matched flight ePSF render (control)}
D1 & $36{,}131$ & $-0.23 \pm 0.05$ ($-4.9$) & $-0.71 \pm 0.04$ ($-16.1$) & $-0.04 \pm 0.04$ ($-1.0$) \\
D2 & $36{,}126$ & $-0.36 \pm 0.05$ ($-7.5$) & $-0.65 \pm 0.05$ ($-14.0$) & $+0.01 \pm 0.04$ ($+0.1$) \\
D3 & $36{,}115$ & $-0.01 \pm 0.05$ ($-0.3$) & $-0.41 \pm 0.05$ ($-8.8$) & $+0.05 \pm 0.04$ ($+1.3$) \\
D4 & $36{,}022$ & $+0.10 \pm 0.05$ ($+2.1$) & $-0.77 \pm 0.05$ ($-17.0$) & $-0.11 \pm 0.04$ ($-2.9$) \\
D5 & $36{,}033$ & $+0.07 \pm 0.05$ ($+1.4$) & $-0.31 \pm 0.04$ ($-7.0$) & $-0.02 \pm 0.04$ ($-0.5$) \\
D6 & $36{,}011$ & $+0.22 \pm 0.04$ ($+5.1$) & $-0.15 \pm 0.04$ ($-3.6$) & $-0.01 \pm 0.04$ ($-0.2$) \\
\enddata
\tablecomments{The coherent response $\mathrm{d}\ln N_{\rm sig}/\mathrm{d}\alpha$ of an astrometric perturbation, in units of $10^{-2}$ inverse native detector pixels, for the three perturbation directions of Section~\ref{sec:corpus:setup}. Uncertainties are the standard error across the $N_{\rm exp}$ exposures of that detector, with Equation~\ref{eq:dlnndet} formed independently per exposure; the parenthesized quantity is the corresponding significance. The random-direction column is expected to be consistent with zero and is used as a null test.}
\end{deluxetable*}

\subsection{\hesp\ response template generation}\label{sec:hesptemplates}

\begin{figure}
  \centering
  \includegraphics[width=\linewidth]{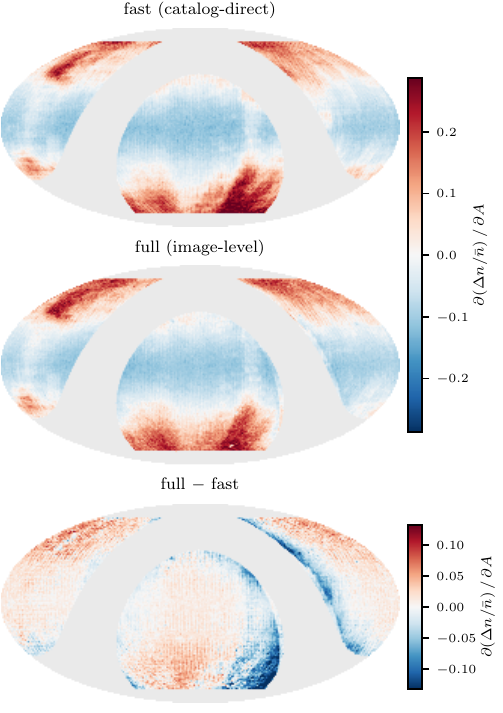}
  \caption{Fast- and full-catalog versions of the same depth response template from Section~\ref{sec:densmodmaps}. The top map shows a 1-year-survey-equivalent version of the depth response template from Figure~\ref{fig:template}, made with the direct-catalog mode using analytic noise approximations, while the center map shows this same quantity computed from the as-flown noise levels and image-level flags from the DR1 first-year dataset, calculated through the full photometric measurement pipeline. Both have $10^7$ sources measured. The bottom map shows the difference between the two, dominated by lost data near the Galactic bulge and varying background/environmental characteristics with respect to the model; this is a reflection of the extent to which the fast catalog approximates the full solution. The maps are smoothed with a 1$^\circ$ Gaussian for visualization.}
  \label{fig:fastvfull}
\end{figure}

The full image-level photometric pipeline of \hesp\ may also be composed with the binning and redshift assignment of Section~\ref{sec:densmodmaps} to produce downstream response templates which incorporate all pixel-level variance characteristics and flags, which we may compare against the direct-catalog templates from \eosp. Figure~\ref{fig:fastvfull} shows a comparison between an observation-equivalent fast \eosp\ catalog and one generated using the full pixel-level variance layers and flags of \hesp. The primary differences come in the form of measurements lost to elevated transient rates towards the Galactic bulge, and varying background or environmental characteristics not captured by the \eosp\ noise model. The \hesp\ version includes all go-backs, unscheduled observations, or downtime, and characterizes the difference in local background conditions that may not be included in the analytic model. Both templates render $10^7$ sources observed in $\sim 10^{6}$ exposures, and we collapse over radial bins for simplicity and to lessen the pixel/voxel sampling requirements. The correlation between the two templates is $r = +0.940$, with the difference's RMS around 1/3 of either template's RMS. The fast \eosp\ template at this density was generated in $0.53$ GH200 GPU-hours, versus 19 4$\times$GB200 node-hours (76 Blackwell GPU-hours) for the full \hesp\ template (the generational differences being modest [$\sim10\%$] for fp64 workloads).

\subsection{Astrometric error marginalization}\label{sec:astmarg}

\begin{figure}
  \centering
  \includegraphics[width=\linewidth]{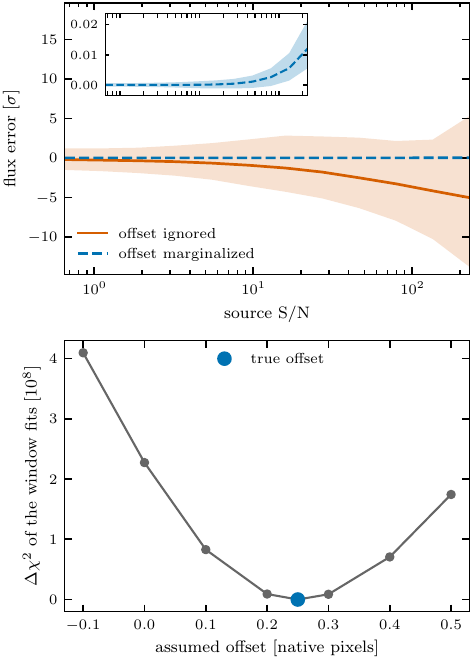}
  \caption{Marginalizing the effect of a coherent position offset. A uniform 0.25-pixel astrometric shift along detector-$x$ is applied across every source in the reference exposure of Section~\ref{sec:hesperos:valtractor}, and fit assuming no shift. This is chosen to be deliberately dramatic to highlight the effect. Sources are rendered at their \textsc{Tractor}-measured real-image fluxes. \textbf{Top}: the error with respect to the fit at the true position in units of sigma as a function of source input S/N, for the cases where the offset is ignored and allowed to affect the measurement without treatment (orange), or if it is marginalized with \hesp\ (blue dashed). The 16th--84th percentiles are shown as shaded bands; the inset shows the marginalized case on an expanded axis range to make these visible. \textbf{Bottom}: the shift in $\chi^2$ of the windowed fits for each assumed offset in the fit, minimized at the true 0.25-pixel offset. A grid scan is used for illustration, though the gradient is used for the optimization in practice.}
  \label{fig:offset-marg}
\end{figure}

To demonstrate another example of how the \hesp\ gradients may be used practically, we assess a 0.25-pixel coherent detector-$x$ offset in the reference exposure fluxes, rendering with the \textsc{Tractor} measured fluxes at the offset location but performing the fit at the original catalog location. Figure~\ref{fig:offset-marg} shows the effect on the flux recovery as a function of original source signal-to-noise: suffering the effect of the offset blindly would incur a significant bias for bright sources, while marginalizing over the shift even with a uniform prior is able to recover unbiased fluxes with even the 16th--84th percentiles lying within $0.02\sigma$ of zero at all S/N. The effect is easily constrained by the fit due to the large number of sources sharing the coherent offset, though other population or S/N-dependent effects may merit more selectivity or different approaches. The marginalization required six derivative passes and a single second derivative, in total equivalent to about eleven times the work of a single photometric fit on the image. This single parameter is a deliberately simple case, but the comparative cheap scalings of additional directions and the overall speed of the measurement operator highlight the potential for more sophisticated treatments of more complex or wider varieties of effect.

\section{Discussion}\label{sec:discussion}

\subsection{Implications of increased throughput}\label{sec:discussion:throughput}

The GPU-accelerated throughput of \hae\ fundamentally changes the calculus around large-scale reprocessing campaigns, with the processing of small regions achievable in a matter of hours and survey-scale runs in a few days on a single node. Systematic templates may be estimated directly and informed by measurement-level models of the coupling as per Section~\ref{sec:densmodmaps:maps}, processed downstream \emph{en masse}. Not limited by Monte Carlo sampling in the generation of these response templates, the only variable that sets the computing requirement is the desired angular and radial resolution, with realistic template generation taking of order a few hours at production resolution on a single GH200 node.

The derivative is also now available through the pipeline at scale, enabling gradient-based optimization of various steps and enabling new classes of inference. The ability to rapidly iterate on the measurement configuration (for instance, the background, PSF, or zodiacal emission models), and rapidly react to incoming instrument data as the survey progresses, enables rapid iteration and optimization of the analysis pipeline and understanding of the instrument response that was not previously practically (or financially) possible.

\subsection{Relationship to source injection}\label{sec:discussion:injection}

A derivative-obtained pipeline response naturally synergizes with source injection, and indeed provides a powerful framework on which to leverage it. Injections sample all (potentially non-differentiable) nonlinearities and discrete aspects of a pipeline, such as detailed deblending topology, image-level transients and artifacts, catalog cuts, and masking decisions. The cost of additional gradients is typically cheap with respect to the cost of a full rerun of the measurement operator (Section~\ref{sec:composition:diff}), providing inexpensive additional information per injection, and the speed of the production measurement itself has now entered a regime which enables dense, many-realization injections across large sky areas.

The responses obtained with this framework allow for characterization of or marginalization over just a class of effects which synthetic source injection is forced to inherit \emph{ab initio}, namely, the response due to features like the detector gain calibration, astrometric calibration, or PSF.\ Detailed characterization of these effects at an instrument level is still critical to obtain the best possible measurements, but the differentiable approach presented here provides a mechanism by which to either assess the downstream impact or absorb a lack of knowledge into the error budget.

Derivatives are appropriate for obtaining the pipeline response directly in regimes and along directions where that response is smooth, while highly nonlinear effects are better served by additional injections. A full injection pipeline also requires extensive bookkeeping and records of both production measurement configuration and input truth, of which the rendering and measurement substrates are just components. The \eosp\ renderer may be readily replaced with an alternative such as the Sky Simulator or \galsim\ (with appropriate considerations for the LVF response), still making use of the differentiable measurement operator \hesp, or of the downstream binning, redshift estimation, and tomographic assignment (which, though implemented in \eosp, does not oblige one to use its renderer).\ \hesp\ and \eosp\ represent the differentiable base upon which a companion paper~(S. Everett et al. 2026, in preparation) will develop the injection application. 

\subsection{Limitations}\label{sec:discussion:limitations}

We note some important limitations of the capabilities demonstrated here.

The scope shown here is solely differential, with a deliberately unrealistic, flat-in-redshift input distribution intended only to measure the response of the pipeline. As mentioned in Section~\ref{sec:composition:diff}, the derivatives may be suspect in cases where the response is strongly nonlinear, which for the variance path is also a strong function of the noise regime being considered. Large perturbations that violate the static assumptions (\emph{e.g.}, perturbing the position of a source sufficiently to make it land in a different measurement window, into a degenerate blend, or into a mask) will also invalidate the derivatives, though hard errors are generally raised in the code when such instances are seen.

We also note some of the effects for which the fast direct-catalog mode of Section~\ref{sec:eosphoros:rendering} is insufficient, and must leverage the full composition of \hae. Section~\ref{sec:corpus} itself presents one of these, the coupled astrometric PSF/LVF model fidelity and response. Other such effects include realistic background variation and environmental dependence; ePSF zone and window topology; and image-level flags. The survey hit pattern used in Section~\ref{sec:densmodmaps} is an approximation based on the full mission plan; this is a good approximation at large scales when aggregating an ensemble of measurements over a wide period in time, but will become increasingly inaccurate at finer spatial and temporal scales due to real-survey features such as flagged or go-back observations.

The normalization of Section~\ref{sec:corpus} is dependent on the signal-to-noise distribution, which may be varied as appropriate for the situation.

\section{Conclusions}\label{sec:conclusions}

We have presented a batched, GPU-accelerated, differentiable forced-photometry operator which reproduces the SPHEREx measurement, \hesp, and a differentiable spectrophotometric forward model that renders sources into frames, \eosp, and demonstrated how their composition may be used to investigate the measurement response to an astrometric perturbation and to the assumed PSF model. By propagating gradients through the full measurement pipeline, we may directly measure the response of the downstream quantities to a given systematic perturbation.

The \hesp\ measurement pipeline is implemented as a Cholesky decomposition which casts an entire exposure as a set of independent linear systems and efficiently batches them for JAX compilation and deployment on GPUs. Validated per-source against the public \textsc{Tractor} fork and at catalog-level against a production catalog (median deviation $0.006\sigma$ over $7.2 \times 10^{8}$ mutually unflagged measurements), the dramatic increase in speed, $8.5 \times 10^4$ exposures per hour on a four-GPU GB200 node, $\sim700\times$ the throughput per node-hour relative to the CPU-based \textsc{Tractor} pipeline, enables a range of new capabilities.\ \hesp\ is currently being evaluated for adoption as the production SPHEREx photometry pipeline for releases following DR1.

The forward model \eosp\ is validated against \galsim\ to less than a few parts in $10^6$ of the stamp peak for a variety of profiles, and against the SPHEREx Sky Simulator. The same downstream derivative propagation that captures the response of the measurement operator to a systematic may also be fed directly by the forward model, producing a direct estimate of the downstream impact of a modeled exposure-level systematic. The speed of this operator enables this at scale, processing a full survey ($\mathcal{O}(10^{10})$ photometric measurements, $\mathcal{O}(10^8)$ sources) with redshift estimation and the derivative in of order a single GPU-day. The forward model and the measurement operator may be easily composed to directly measure selection responses to various effects, flags, or measurement choices, or used independently. An initial demonstration of this capability strongly detected ($195$--$300\sigma$ across detectors,\ Table~\ref{tab:response}) a mismatch between a scene rendered using a pre-flight ray-traced PSF model and one measured with the flight data-informed effective PSF model.

Accelerated photometry processing pipelines allow for a rapid rate of iteration and optimization of the measurement, and naturally benefit efforts to measure the selection function of a survey through source injection, or to train emulators of injections. Differentiability may naturally compound the utility of injections by allowing many additional perturbations to be measured per injection at an amortized cost, or for a cost independent of the number of parameters with reverse-mode differentiation. The intrinsic speed of \hesp\ makes the injection application a natural one, as they typically rely on rerunning the production pipeline and would benefit from being able to do so a large number of times; this will be the subject of a companion paper~(S. Everett et al. 2026, in preparation). Differentiability also enables the use of Hamiltonian Monte Carlo or other gradient-based sampling techniques, offering new possibilities for Bayesian treatments of systematics. For challenging large-scale cosmological targets such as $\fnl$, careful characterization of the selection function is essential, and increases in scalability and capability as shown here are a step on the path to achieving that goal.

\section*{Data and code availability}

\hae\ are under ongoing internal development as part of SPHEREx's Level 4 science mission and will be made public at a later date along with the catalog products, per the mission's public schedule and following the public release of the SPHEREx Sky Simulator. Level 2 and Level 3 data products were from internal releases of the DR1 SPHEREx pipeline which have since been made available through the NASA/IPAC InfraRed Science Archive (IRSA)\footnote{\url{https://irsa.ipac.caltech.edu/Missions/spherex.html}}, following the mission's public schedule~\citep{https://doi.org/10.26131/irsa678}. The comparison of Section~\ref{sec:hesperos:valtractor} is done against the public \texttt{tractor-spherex} fork at commit \texttt{066680f}.

\begin{acknowledgments}
  We acknowledge support from the SPHEREx project under a contract from the NASA/Goddard Space Flight Center to the California Institute of Technology. JRC wishes to acknowledge support from the David \& Ellen Lee Fellowship. The High Performance Computing resources used in this investigation were provided by funding from the JPL Information and Technology Solutions Directorate. Computations were run on the \textsc{Stampede3}, \textsc{Lonestar6}, and \textsc{Vista} clusters provided by the Texas Advanced Computing Center (TACC)\footnote{\url{http://www.tacc.utexas.edu/}}. Part of this work was done at Jet Propulsion Laboratory, California Institute of Technology, under a contract with the National Aeronautics and Space Administration (80NM0018D0004). The authors would also like to acknowledge the use of Anthropic's Claude \textsc{Opus} 4.8, 5.0, and \textsc{Fable} 5 models for assistance in literature review, code implementation, manuscript proofreading, and refinement of some figures. All generated content was assessed by the authors who take full responsibility for the content presented herein.
\end{acknowledgments}

\software{
  \tractor~\citep{lang2016tractor},
  \texttt{tractor-spherex} (\url{https://github.com/Caltech-IPAC/tractor-spherex}),
  The SPHEREx Sky Simulator~\citep{crill2025},
  JAX~\citep{jax2018github},
  \texttt{numpy}~\citep{harris2020},
  Claude \textsc{Opus} 4.8, 5.0, and \textsc{Fable} 5 (\url{https://www.anthropic.com/})
}

\bibliography{hesperos_eosphoros}
\bibliographystyle{aasjournalv7}

\appendix
\section{\hesp\ validation details}
\label{ap:hespval}

\begin{figure*}
  \centering
  \includegraphics[width=\textwidth]{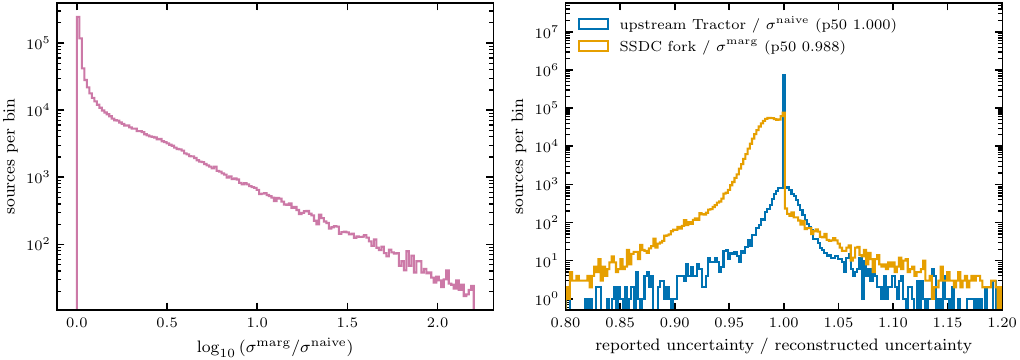}
  \caption{Uncertainty conventions and comparison for the reference exposure. \textbf{Left}: The ratio of the \hesp\ marginal to na\"{\i}ve errors, illustrating the effects of covariant (blended) contributions. \textbf{Right}: A comparison of \hesp's uncertainty conventions to the upstream \tractor\ code (blue), which provides the na\"{\i}ve error $\sigma^{\rm naive}$, and to the public SSDC fork (orange) which incorporates the blending terms into the marginal uncertainty $\sigma^{\rm marg}$. The SSDC fork's seeming 1\% deficit/pileup is a result of their choice to freeze the background term $b$ in the calculation of the uncertainty, which we choose to marginalize; freezing it in \hesp\ recovers a similar degree of agreement as to the upstream branch's na\"{\i}ve uncertainty.}\label{fig:errconv}
\end{figure*}

For the flux measurements themselves, the SSDC fork and the upstream \textsc{Tractor} code were found to give identical results, the differences being in the uncertainties. Figure~\ref{fig:errconv} shows the comparison of $\sigma^{\rm naive}$ and $\sigma^{\rm marg}$ to the upstream \textsc{Tractor} and to the SSDC fork, respectively, for the reference exposure, and a comparison of \hesp-derived $\sigma^{\rm naive}$ and $\sigma^{\rm marg}$. $\sigma^{\rm naive}$ reproduces the upstream build's uncertainty to $1.000$ from the first to the ninety-ninth percentiles, while $\sigma^{\rm marg}$ matches the \tractor\ fork to a median $0.988$ over the first to ninety-ninth percentiles (the difference being due to how the reference's Fisher matrix fixes its background term).

\begin{figure*}
  \centering
  \includegraphics[width=0.7\textwidth]{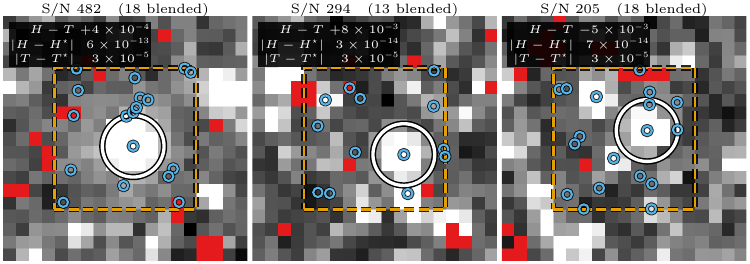}
  \caption{A sampling of well-agreeing measurements in the reference exposure between the two implementations for some high signal-to-noise ($S/N > 200$) sources with low flag rates. The measured source is highlighted with a white circle, while all photometered sources in the unique cutout (orange dashed square) are shown as blue circles. The inset shows the \hesp-\textsc{Tractor} difference for that source (``$\mathrm{H}-\mathrm{T}$'') in units of the \textsc{Tractor} measurement's $\sigma^{\rm marg}$, as well as each flux's absolute distance from its own WLS optimum (denoted $H^\star$ and $T^\star$ for \hesp\ and \tractor, respectively) in units of its own $\sigma^{\rm marg}$. Red pixels are flagged in the L2 images and excluded from the fit.}\label{fig:cutouts_clean}
\end{figure*}

\begin{figure*}
  \centering
  \includegraphics[width=0.7\textwidth]{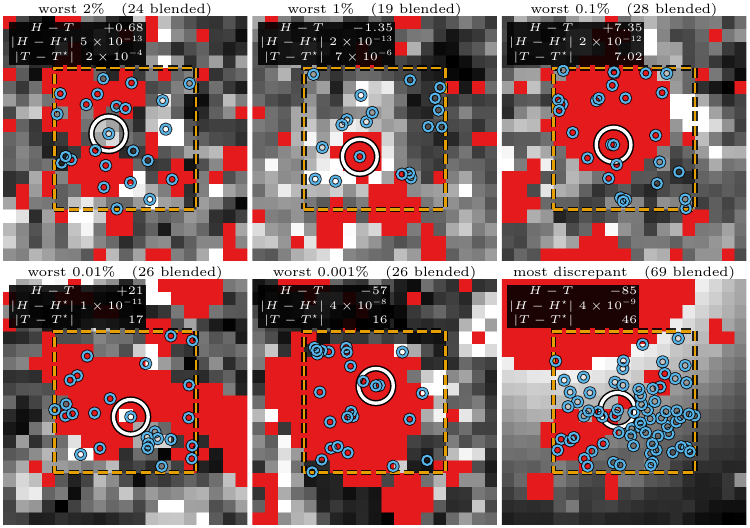}
  \caption{A sampling of particularly discrepant measurements between \hesp\ and the \textsc{Tractor} reference, at intentionally distant locations in the reference exposure. Individual flux measurements whose flux differences are closest to a few tiered percentiles are shown, highlighted with white circles, along with their $11 \times 11$-pixel measurement windows and the other simultaneously-photometered sources shown as blue circles. Percentiles from the worst 2\% through to the worst 0.001\% are shown, as well as the single most discrepant measurement in the exposure. Inset values are as in Figure~\ref{fig:cutouts_clean}, and as before red pixels denote flagged pixels excluded from the fit. The most-discrepant cases are generally driven by excessive flagging, but even in this regime \hesp\ is consistently closer to its own reference optimum ($H^\star$). The single most discrepant case is driven by a large number of potentially spurious or obscured sources near the halo of a bright star off the upper-left edge of the image cutout.}\label{fig:cutouts}
\end{figure*}

Some well-agreeing measurements for bright sources in this dense Galactic reference exposure field are shown in Figure~\ref{fig:cutouts_clean}. Figure~\ref{fig:cutouts}, on the other hand, shows a sampling of the most discrepant cases between the two implementations for the reference exposure. These cases are driven by degenerate blends and large numbers of flagged pixels which are excluded from the fit, making the fits poorly conditioned (these measurements, or even the entire exposure, would likely be flagged downstream in a real analysis). The \hesp\ solution is consistently closer to its WLS optimum even for the worst cases.

Twenty exposures over a range of observing densities were used to compare the implementations, checked against a set of conservative predefined criteria: the recovered flags must be exact; there must be $< 0.75\%$ of sources lying beyond $1\sigma$ of the (exact version of the) reference solution; and the distance between the solvers must correlate with the reference's own residual at $\rho \geq 0.9$. 15 of 20 exposures pass all of these criteria. Of the five that do not, three are heavily masked frames that would be cut and flagged to the survey to re-try (typically due to transient events); one was the densest field in the comparison with $39\%$ degenerate windows; and the last was a single flag difference in a single row due to a boundary effect. For a matched set of un-flagged observations, \hesp\ is closer to the exact solution for 20/20 exposures, as assessed by the fraction of sources beyond $1\sigma$ from the optimum; this was \emph{e.g.} $42.2\%$ for the reference versus $12.7\%$ for \hesp\ for one of the highly degenerate exposures. On an identical set of sources with identical cuts, the \hesp\ direct solution outperforms the iterative reference by this metric for all situations which were considered.

\section{S\'ersic rendering validation}
\label{ap:gsval}

To assess the extended S\'ersic rendering of \eosp\ against that of \galsim, we compare the measured flux differences for a random population of $9{,}799$ rendered Legacy Survey shapes using the two codes. We use \galsim\ to draw S\'ersic profiles convolved with a Gaussian PSF at the centers of $18\times18$ oversampled subpixel cells on the same periodic canvas used by \eosp, placed with the same sub-pixel weights, and summed to SPHEREx native pixels.\ \eosp\ renders the source using its default settings. Because the PSF is shared, this tests the profile, shear and orientation, cell integration, and placement. Shapes are drawn jointly in (index, half-light radius, axis ratio, and position angle) from the $1.45\times10^{6}$ extended sources ($r_e \geq 1''$) of twelve Reference Catalog fields. The fractional flux difference $|\delta f /f|$ is computed using least-squares fits of the \galsim\ profile to both the \galsim\ and \eosp\ images, with uniform pixel noise.

\begin{figure}
  \centering
  \includegraphics[width=0.6\textwidth]{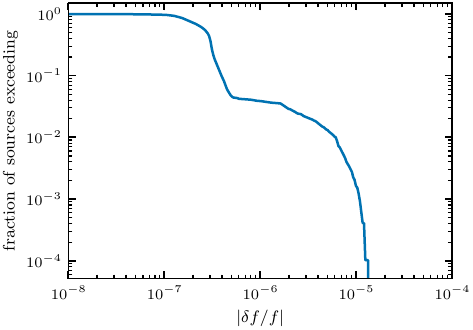}
  \caption{The fraction of sources exceeding a fractional flux difference $|\delta f/f|$, comparing the \eosp\ to the \galsim\ extended renders for a sample of $9{,}799$ galaxies drawn from the Reference Catalog (Legacy Survey) shape population at random sub-pixel positions, image rotations, and parities. The flux is measured using the same (\galsim) profile for each image. 90\% of galaxies beyond $5\times10^{-7}$ are ones that are too extended to fully fit in the canvas, which are only expected to be accurate to about this level.}
  \label{fig:gsval}
\end{figure}

Figure~\ref{fig:gsval} shows the fraction of sources exceeding a fractional flux difference $|\delta f/f|$, with values small throughout; of the rendered galaxies disagreeing beyond $5\times10^{-7}$, 90\% of these are very large sources which extend beyond the largest available canvas edge. These are expected to only be accurate at around this level. The median flux difference is $3\times10^{-7}$ and the largest fractional difference was $1.4\times10^{-5}$. These differences are far subdominant to the measurement noise, with even that largest fractional error applied to the brightest un-saturated source measured in the reference exposure of Section~\ref{sec:hesperos:valtractor} (S/N $514$) resulting in a flux difference of $0.007\sigma$, with a median difference $1.5\times10^{-4}\sigma$.

\end{document}